\documentclass[fleqn,usenatbib]{mnras}

\usepackage{newtxtext,newtxmath}

\usepackage[T1]{fontenc}
\usepackage{ae,aecompl}

\usepackage{graphicx}	
\usepackage{amsmath}	
\usepackage{amssymb}	

\newcommand{\Msun}{\rm{M}_{\odot}}

\usepackage{amsmath}
\DeclareMathOperator{\sech}{sech}

\title[Baryonic effects on the Pal 5 stream]{Effects of baryonic and dark matter substructure on the Pal 5 stream}

\author[Banik \& Bovy]{
Nilanjan Banik,$^{1,2}$\thanks{E-mail: banik@lorentz.leidenuniv.nl}
Jo Bovy$^{3,4}$
\\
$^{1}$GRAPPA Institute, Institute for Theoretical Physics Amsterdam 
and Delta Institute for Theoretical Physics, University of Amsterdam, \\
Science Park 904, 1098 XH Amsterdam, The Netherlands\\
$^{2}$Lorentz Institute, Leiden University, Niels Bohrweg 2,
Leiden, NL-2333 CA, The Netherlands\\
$^{3}$Department of Astronomy and Astrophysics, University of Toronto, 
50 St. George Street, Toronto, ON, M5S 3H4, Canada\\
$^{4}$Alfred P. Sloan Fellow
}

\date{Accepted XXX. Received YYY; in original form ZZZ}

\pubyear{2018}

\begin{document}
\label{firstpage}
\pagerange{\pageref{firstpage}--\pageref{lastpage}}
\maketitle

\begin{abstract}
Gravitational encounters between small-scale dark matter substructure and cold stellar streams in the Milky Way halo lead to density perturbations in the latter, making streams an effective probe for detecting dark matter substructure. The Pal 5 stream is one such system for which we have some of the best data. However, Pal 5 orbits close to the center of the Milky Way and has passed through the Galactic disk many times, where its structure can be perturbed by baryonic structures such as the Galactic bar and giant molecular clouds (GMCs). In order to understand how these baryonic structures affect Pal 5's density, we present a detailed study of the effects of the Galactic bar, spiral structure, GMCs, and globular clusters on the Pal 5 stream. We estimate the effect of each perturber on the stream density by computing its power spectrum and comparing it to the power induced by a CDM-like population of dark matter subhalos. We find that the bar and GMCs can each individually create power that is comparable to the observed power on large scales, leaving little room for dark matter substructure, while spirals are subdominant on all scales. On degree scales, the power induced by the bar is small, but GMCs create small-scale density variations that are similar in amplitude to the dark-matter induced variations but otherwise indistinguishable from it. These results demonstrate that Pal 5 is a poor system for constraining the dark matter substructure fraction and that observing streams further out in the halo will be necessary to confidently detect dark matter subhalos.
\end{abstract}




\section{Introduction}

A crucial prediction of the $\Lambda$CDM framework is the presence of a large number of subhalos orbiting within a Milky Way sized host halo  \citep{Klypin1999,Moore1999,Diemand2008,Springel2008}. Detecting these subhalos would not only prove that dark matter is a form of matter capable of clustering on sub-galactic scales, but would also give crucial insight into its particle nature and interactions. One purely gravitational method for detecting dark matter substructures is gravitational lensing \citep[e.g.,][]{Mao1997,Dalal2002,Vegetti2012}. Gravitational lensing can allow us to measure the abundance of low-mass substructures around external galaxies \citep{Hezaveh2016,Daylan2017}.

An alternate but equally promising purely gravitational method for detecting these subhalos is to use cold stellar streams that originate as a globular cluster falls into our Galaxy's gravitational potential and gets tidally disrupted. The density of such a stream is largely uniform along its length in the absence of perturbations. A gravitational encounter with a dark matter subhalo perturbs the stream density resulting in gaps in the density \citep{Ibata2001,Johnston2002,Siegal-Gaskins2008,Carlberg2009}. Much work has been done in the last few years towards modeling and analyzing these gaps and inferring the properties of the subhalos that the stream encountered \citep[e.g.,][]{Yoon2011,Carlberg2012,Carlberg2013,Erkal2015,Erkal2015a,Sanders2016}.

Recently, a statistical approach for inferring the properties of the dark matter subhalos using the stream density power spectrum and bispectrum was proposed by \citet{Bovy2016a}. Applying this approach to data on the density of the Pal 5 stream from \citet{Ibata2015}, the authors computed the observed power spectrum of the stream density and by matching this to simulations used this to constrain the number of cold dark matter subhalos orbiting within the Galactic volume of the Pal 5 orbit. In doing so however, the authors neglected any effects from the baryonic perturbers in the Galaxy such as the central bar, the spiral structure, giant molecular clouds (GMCs), and the globular cluster (GC) system. Because of this neglect, they pointed out that their measurement of the number of dark matter subhalos was in fact a robust upper limit to the amount of dark matter substructure. The effect of the bar on stellar streams orbiting near the center of the Galaxy has been shown to be potentially large \citep[e.g.,][]{Hattori2015,Erkal2017,Pearson2017}, especially for the Pal 5 stream because it is in a prograde orbit with respect to the disk and everything orbiting within it. Therefore the density of the Pal 5 stream can be affected by the Galactic bar \citep{Pearson2017,Erkal2017}, GMCs \citep{Amorisco2016}, and likely spiral structure as well. All these findings then beg the question: is the Pal 5 stream a good probe for detecting dark matter substructures in our Galaxy? To answer this question, in this paper we perform a detailed investigation of the possible baryonic perturbers individually, using up-to-date constraints on their properties, and we compute the effect each one has on the Pal 5 stream.

The paper is structured as follows: In Section \ref{sec:pal5}, we introduce the Pal 5 stream, observations of its density, and a brief description of the CDM subhalo model used for the stream-subhalo encounters. In Section \ref{sec:bar}, we discuss the bar model setup and decide on the intervals over which the bar model parameters will be varied. In the subsection  \ref{sec:bar_effect} we discuss how we model the effects of the bar on the Pal 5 stream density, followed by subsection \ref{sec:bar_result}, where we present the results of the mock Pal 5 stream's power spectrum as a result of varying the bar models. Next, in Section \ref{sec:spiral}, we describe the model of the spiral potential and present the results of varying the spiral arm potential's model parameters in subsection \ref{sec:spiral_effect}. In Section \ref{sec:GMC}, we discuss how the GMCs are included in our simulations and in subsection \ref{sec:Pal5_peri}, we explore how their effect on Pal 5 stream's density changes on varying Pal 5's pericenter within a range that is allowed by observations. We present the results of the GMC impacts on Pal 5 stream in subsection \ref{sec:GMC_results}. Then in Section \ref{sec:GCs}, we describe how we incorporated the Galactic population of GCs in the stream simulations and discuss the results. Finally, in Section \ref{sec:conclusion}, we discuss all the results and present our conclusions.

All of our modeling is done using tools available as part of the \texttt{galpy} galactic dynamics Python package\footnote{Available at \url{https://github.com/jobovy/galpy}~.} \citep{Bovy2015}. 

\section{The Pal 5 stream}\label{sec:pal5}

The Pal 5 stream is a cold stellar stream emanating from its namesake progenitor, the Pal 5 globular cluster. It was discovered by \citet{Odenkirchen2001} using data from the Sloan Digital Sky Survey (SDSS) \citep{York2000}. Its trailing arm spans over $\sim 14 ^{\circ}$ while its leading arm is only around $\sim 8^{\circ}$ \citep{Bernard2016}. Since its discovery, there has been a number of follow up studies to measure its stellar density \citep[e.g.,][]{Odenkirchen2003,Carlberg2012a,Ibata2015}. In what follows, we briefly describe how we model the smooth Pal 5 stream in this paper. 

Following \citet{Bovy2014}, we generate a mock Pal 5 stream using a frequency-angle ($\Omega,\theta$) framework in the \texttt{MWPotential2014} \citep{Bovy2015}. This method requires the phase space coordinates of the progenitor, the velocity dispersion $\sigma_{v}$ of the stars and the time $t_{d}$ since disruption commenced. Following \citet{Fritz2015}, we set the phase space coordinates of the Pal 5 globular cluster to $({\rm RA,Dec},D,\mu_{\alpha}\cos\delta,\mu_{\delta},V_{\rm{los}}) = (229^{\circ}.018,-0^{\circ}.124,23.2 {\rm ~ kpc},-2.296 {\rm ~mas~yr^{-1}},-2.257 {\rm ~mas~yr^{-1}}, \\
-58.7 {\rm ~ km ~ s^{-1}})$. Following \citet{Bovy2016a}, we set $\sigma_{v} = 0.5$ km/s and $t_{d} = 5$ Gyr, because they demonstrated that this gives a close match to all of the data on the Pal 5 stream. The stream generated in the ($\Omega,\theta$) space is then transformed to rectangular Galactocentric coordinates using the approach of \citet{Bovy2014} and from there to the custom $(\xi,\eta)$ stream coordinates defined by \citet{Ibata2015}. For the rest of this paper we will focus only on the trailing arm of the stream in the range $0.65^{\circ} < \xi < 14.35^{\circ}$, because this is the part of the stream for which the best density data exists and it is the part studied in detail by \citet{Bovy2016a}.

Throughout this paper, we compare the effect of baryonic perturbers on the Pal 5 stream to that expected from dark matter subhalos. The modeling of the subhalo population and how it affects the Pal 5 stream is discussed in detail by \citet{Bovy2016a}. Here we briefly describe important aspects of this modeling that are relevant for the discussion in the subsequent sections. Following the approach in \citet{Bovy2016a}, we use the CDM subhalo mass function, $dn/dM \propto M^{-2}$ and model subhalos as Hernquist spheres whose scale radius depends on the subhalo mass according to the fitting relation $r_{s}(M)=1.05 {~\rm kpc}~(M/10^{8} \Msun)^{0.5}$; this relation was obtained by fitting Hernquist profiles to the circular velocity-$M$ relation from Via Lactea II simulations \citep{Diemand2008}. The amplitude of the subhalo mass function in the range $10^5$ to $10^9\,\rm{M}_\odot$ is determined from the number of dark matter subhalos within 25 kpc in the Via Lactea II simulations. Because the number of subhalos is based on a dark-matter-only simulation, this number does not take into account the possible disruption of some fraction of the subhalo population in the inner Galaxy due to tidal shocking by the disk and bulge, which in simulations leads to a factor of two to four lower subhalo abundance \citep[e.g.,][]{DOnghia2010,Sawala2016}.

\section{The effect of the Galactic bar}
\label{sec:bar}

\begin{figure}
\includegraphics[width=0.5\textwidth]{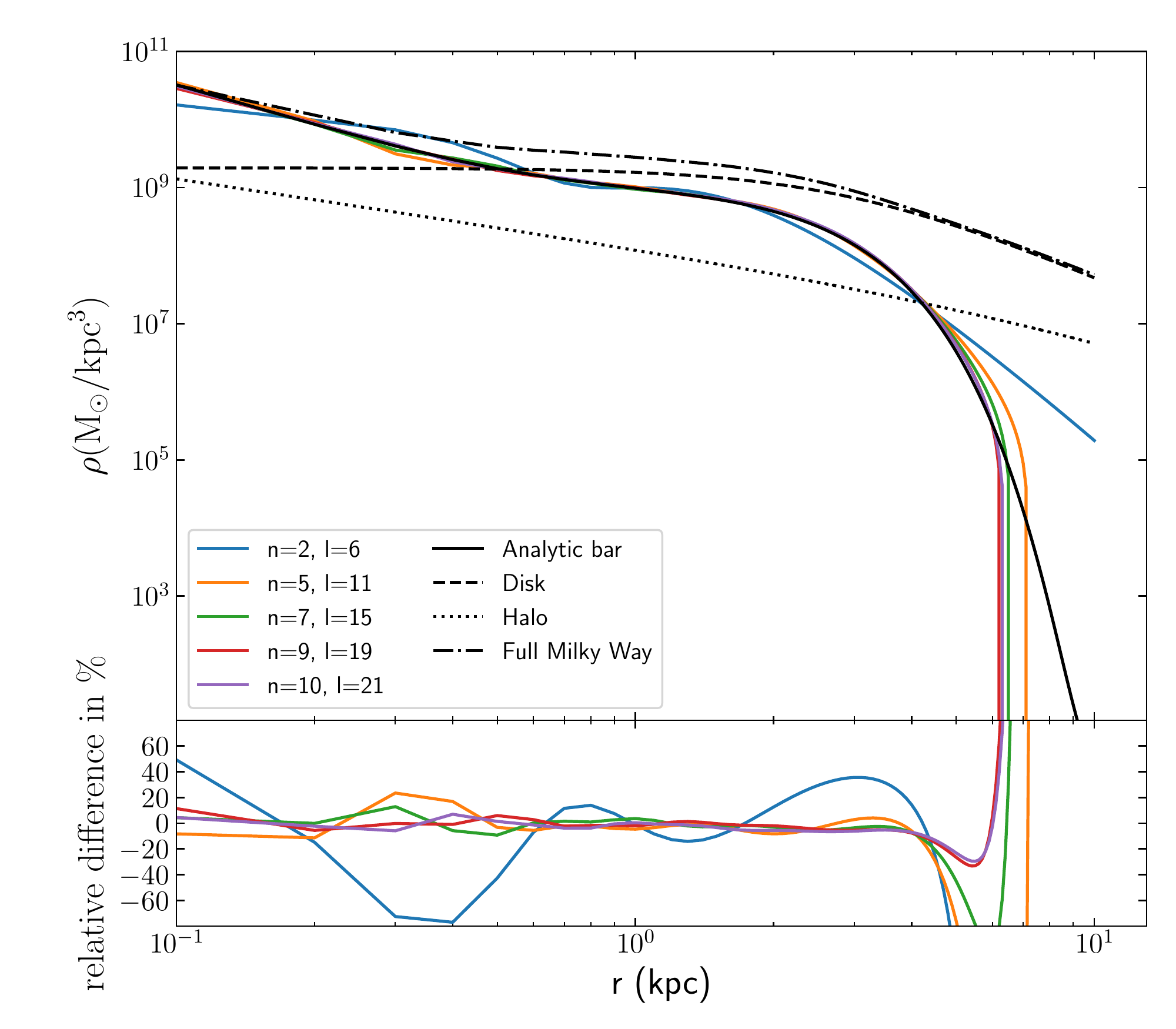}
\caption{Reconstructed density of the bar model for different expansion orders in the basis-function approach to obtaining the potential. The black curve show the analytic density of the bar. The dashed, dotted and dash-dotted curves represent the density for the disk, the halo and the full Milky Way potential respectively, in the \texttt{MWPotential2014} model for the potential that we use. The bottom panel displays the relative difference between each reconstructed density and the analytic density of the bar. The expansion with $n=9$ and $l=19$ gives an excellent match to the input density of the bar model.}
\label{fig:SCF_n_l}
\end{figure}

\subsection{Bar models}

We model the Galactic bar with the triaxial, exponential density profile from \cite{Wang2012}: 

\begin{equation}
\rho_{\rm{bar}} = \rho_{0}\left[\exp(-r_{1}^{2}/2) + r_{2}^{-1.85}\exp(-r_{2})\right].
\label{eq:rhobar}
\end{equation}
Where the functions $r_{1}$ and $r_{2}$ are defined as
\begin{equation}
r_{1} = \left[\left((x/x_{0})^{2} + (y/y_{0})^{2}\right) + (z/z_{0})^{4}\right]^{1/4}
\end{equation}

\begin{equation}
r_{2}=\left[\frac{q^{2}(x^{2} + y^{2}) + z^{2}}{z_{0}^{2}}\right]^{1/2}
\end{equation}

with $x_{0}=1.49$ kpc, $y_{0}=0.58$ kpc and $z_{0}=0.4$ kpc, $q=0.6$ and $\rho_{0}$ is the normalization for a given mass of the bar. To compute the potential from the density, one needs to solve the Poisson equation. We do this by following the basis-function expansion method from \citet{Hernquist1992}, in which we expand the potential and density into a set of orthogonal basis functions of potential-density pairs consisting of spherical harmonics indexed by $(l,m)$ and a radial set of basis functions indexed by $n$. The method requires a single distance scale parameter $r_s$ to be set as well. This method is implemented in \texttt{galpy} and we compute the expansion coefficients by setting the scale length $r_{s} = 1$ kpc. To find the minimum order of expansion coefficients required to get a close reconstruction of the density from the potential, we reconstructed the density for a range of expansion orders and compared the resulting density to the analytic form of the density in Equation \eqref{eq:rhobar}. The colored curves in Figure \ref{fig:SCF_n_l} show the reconstructed density for some of the expansion orders. The bottom panel displays the percentage difference between each case of expansion order and the analytic density (Eq. [\ref{eq:rhobar}]). The departure of the reconstructed density at Galactocentric $r > 5$ kpc does not affect the analysis since at that radial distance the disk's contribution to the density and hence the potential is much more important than that of the bar, as shown in the same figure. From visual inspection, we found that for $n=9$ and $l=19$ we get an excellent reconstruction of the density and therefore we used these values for the rest of the analysis. \citet{Pearson2017} also used the basis function expansion technique to model the bar, however they used expansion order up to $n=2$ and $l=6$. As shown in Figure \ref{fig:SCF_n_l}, this gives a poorer reconstruction of the density.

\begin{figure}
\includegraphics[width=0.5\textwidth]{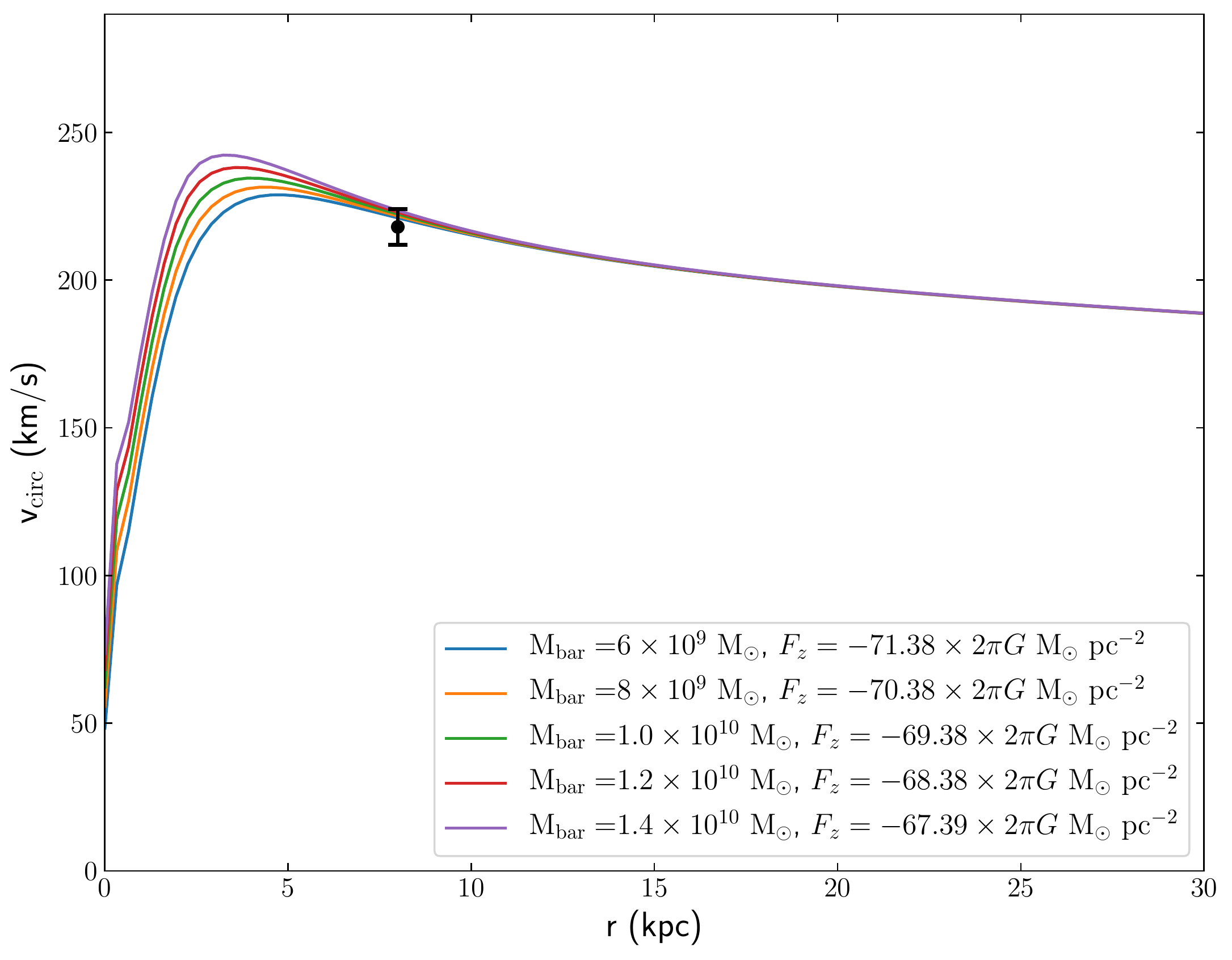}
\caption{Circular velocity curve for models with different bar masses. The black dot with the error bar shows the constraint on the circular velocity at the location of the Sun. The legend of the plot lists the vertical force at $(R,z) = (8,1.1)$ kpc.}
\label{fig:vcirc}
\end{figure}

\begin{figure}
\includegraphics[width=0.5\textwidth]{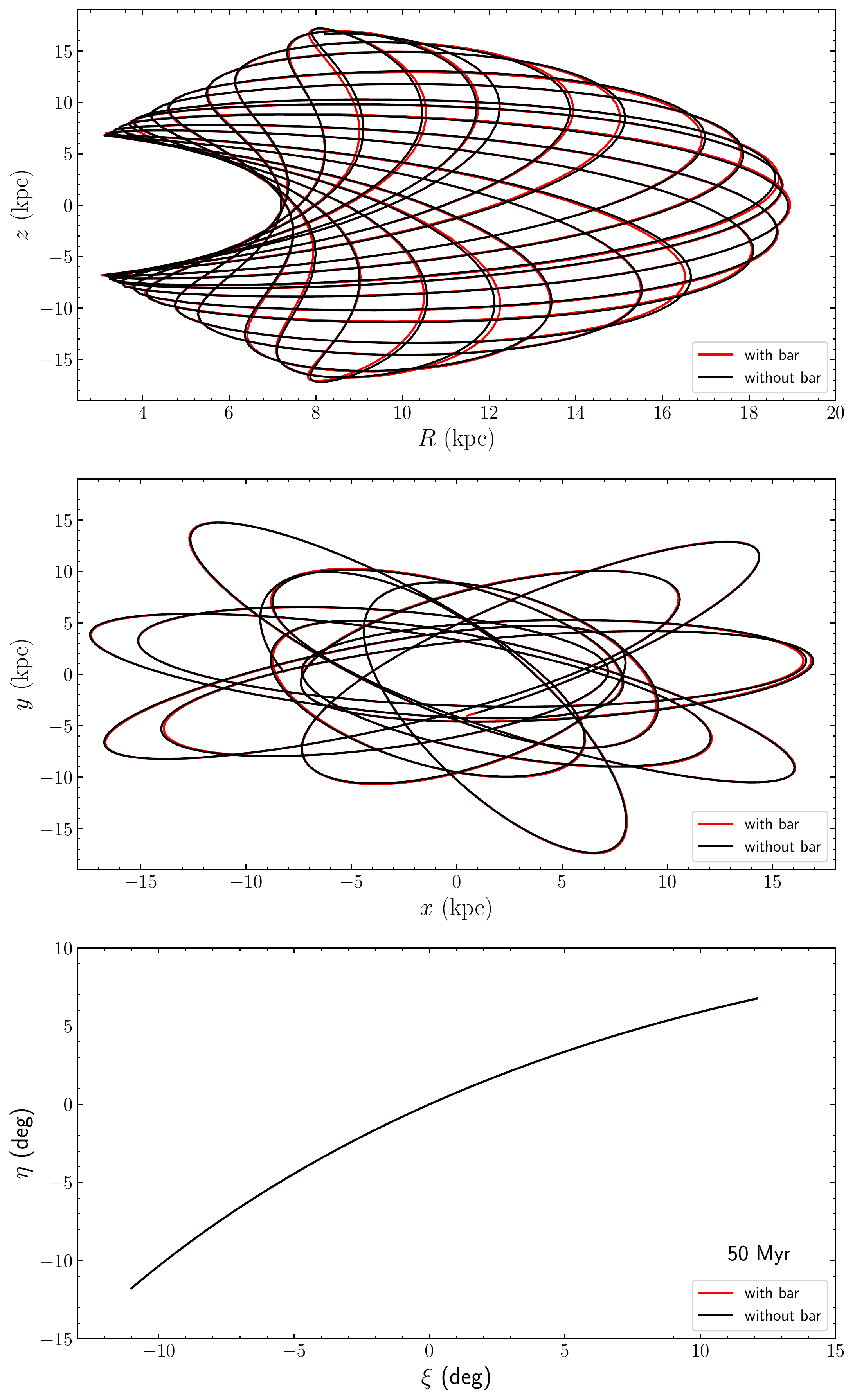}
\caption{Orbit of the Pal 5 globular cluster evolved in the fiducial barred potential (red) versus in the axisymmetric potential (black). The top and the middle panel shows the orbits in Galactocentric $(R,z)$ plane and $(x,y)$ plane for the last 5 Gyr of evolution. The bottom panel compares the orbital evolution in custom $(\xi,\eta)$ coordinates from 50 Myr in the past to 50 Myr in the future, the range which approximately spans the observed Pal 5 stream; there is almost no difference in the present orbit between the barred and axisymmetric potential.}
\label{fig:Pal5_orbit}
\end{figure}

We consider five bar models with masses between $6\times 10^{9}\,\rm{M}_{\odot}$ and $1.4\times 10^{10}]\,\rm{M}_{\odot}$ in increments of $2\times 10^{9}\,\rm{M}_{\odot}$. To incorporate the bar while keeping the same total baryonic mass in our Milky Way mass model, we remove the bulge of mass $5\times 10^{9}\,\rm{M}_{\odot}$ from the \texttt{MWPotential2014} model and any additional mass of the bar above this value is removed from the disk component. Figure \ref{fig:vcirc} shows the resulting circular velocity curve for each model of the bar. It is clear that the circular velocity of all of the models is very similar outside of the bar region and only slightly changes within it. The measurement shown represents the circular velocity constraint of $218 \pm 6 \, \rm{km \ s^{-1}}$ in the solar neighborhood as obtained from APOGEE data by \citet{Bovy2012}. In the same plot, we list the value of the vertical force in each bar model at 1.1 kpc above the plane at this position, which was constrained by \cite{Zhang2013} from the kinematics of K-type dwarfs to be $|F_{z}| = 67 \pm 6 \ (2\pi G \ \rm{M}_{\odot}\ pc^{-2} )$. Therefore, our bar models do not significantly change \texttt{MWPotential2014} outside of the bar region, and our barred models are therefore approximately as good mass models for the Milky Way as \texttt{MWPotential2014}. In addition to varying the mass of the bar, we vary its pattern speed over the grid between the values of $35$ and $61\,\rm{km \ s^{-1} \ kpc ^{-1}}$ in increments of $2\,\rm{km \ s^{-1} \ kpc ^{-1}}$ and we consider ages for the bar between 1 and 5 Gyr in 1 Gyr increments. In each case, the amplitude of the bar is smoothly grown from 0 to its full amplitude following the prescription of \citet{Dehnen2000} over two rotation periods of the bar. As a fiducial model, we consider the model of the bar to be 5 Gyr old, with a mass of $10^{10}\ \rm{M}_{\odot}$, and rotating at a pattern speed of 39 $\rm{km \ s^{-1} kpc^{-1}}$ \citep{Portail2016}. The angle of the bar's major axis with respect to the Sun--Galactic-center line at the present day is in all bar models set to $27^{\circ}$ \citep{Wegg2013}. 

As the non-axisymmetric bar rotates, it imparts kicks to orbiting stars, thereby altering their kinematics. To give an indication of how a barred potential affects an orbiting star, we compare the orbital evolution of the Pal 5 globular cluster in the barred potential to that in an axisymmetric version of the same potential. We construct the latter by setting the expansion coefficients with $m\neq0$ to zero in the basis function expansion. To evolve the orbit of Pal 5 globular cluster, we first integrated it back for 5 Gyr in the past in both the axisymmetric and non-axisymmetric potentials and then integrated it forward to the present in the same potentials. The resulting orbits are plotted in Figure \ref{fig:Pal5_orbit}. In the presence of the bar, the orbit is only slightly altered.

\subsection{Effect of the Galactic bar on the Pal 5 stream}
\label{sec:bar_effect}
 
\begin{figure}
\includegraphics[width=0.5\textwidth]{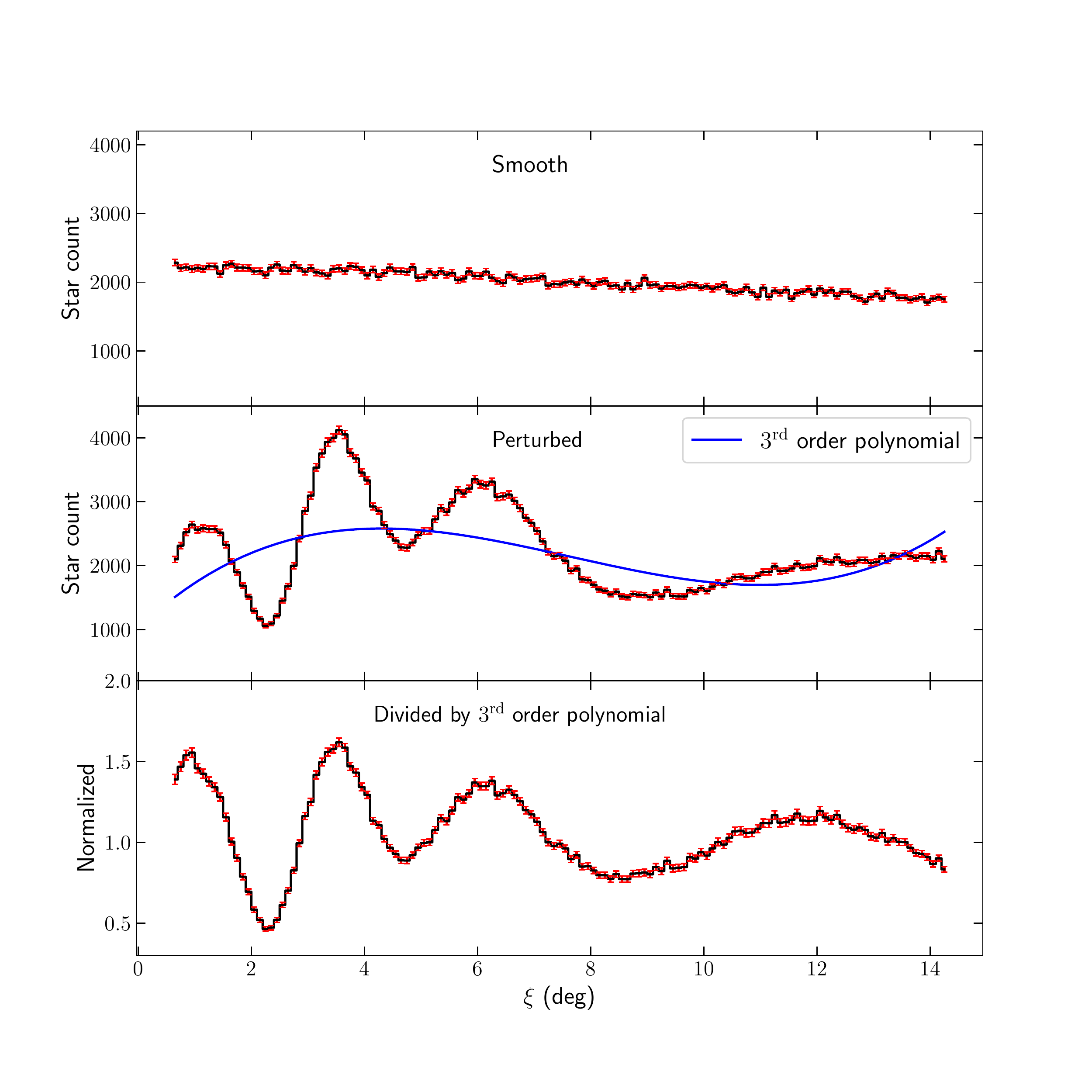}
\caption{Star counts of mock Pal 5 streams. The top panel shows the star count of the stream evolved in the axisymmetric potential. The middle panel shows the star count of the stream evolved in the Milky Way potential with a 5 Gyr old bar of mass $10^{10}\, \rm{M}_{\odot}$ and of pattern speed 39 $\rm{km \ s^{-1} kpc^{-1}}$. The blue curve is the $3^{\rm{rd}}$ order polynomial fit to the star counts with which we normalize the density. The bottom panel shows the perturbed stream star count divided by the polynomial, of which we compute the power spectrum shown in subsequent figures. In each case the sample size is 500,000 and the $\xi$ bin width is $0.1^{\circ}$. The red error bars are the shot noise in each bin.}
\label{fig:nden}
\end{figure}

To compare the density structure of the Pal 5 stream induced by the bar to that due to dark matter subhalos and to the observed density, we evolve a mock Pal 5 stream in the barred Milky Way potentials described above and compute the power spectrum of the stream density. We make use of the same technique as discussed in \citet{Bovy2016a} to compute the power spectrum.  The frequency-angle framework for modeling tidal streams does not support non-axisymmetric potentials. We therefore generate the mock stream in the axisymmetric version of each bar potential and sample from the phase-space coordinates at the present time and the time each star was stripped from the progenitor cluster for many stars. We then integrate each star backwards in time in the same axisymmstric potential to the time of stripping (which is different for each star). Finally, we integrate each star forward in time, now in the barred Milky Way potential until today. This gives the phase-space coordinates of stream stars today due to their evolution in the barred potential. We then transform the coordinates of these stars to $(\xi,\eta)$ coordinates. After selecting all the stars that lie in the observed part of the trailing arm, we bin the sample of stars in $0.1^{\circ}$ bins in the $\xi$ coordinate, to mimic the analysis of \citet{Bovy2016a}. To minimize the shot noise in the density resulting from sampling only a finite number of stars, we sample $\sim$ 500,000 stars. Following the arguments presented in \citet{Bovy2016a}, we normalize the binned density by fitting a third order polynomial to the density and divide the density by this fit. This is done to remove large scale variations in the stream such as could be expected from variations in the stripping rate, which in our analysis is assumed to be constant. We use this normalized density to compute the power spectrum. Figure \ref{fig:nden} shows the star counts of a smooth stream and a stream perturbed by the fiducial bar model. As expected, the unperturbed stream has a largely uniform star count along its length. Evolution in the barred potential results in large density perturbations along the stream.

\begin{figure}
\includegraphics[width=0.5\textwidth]{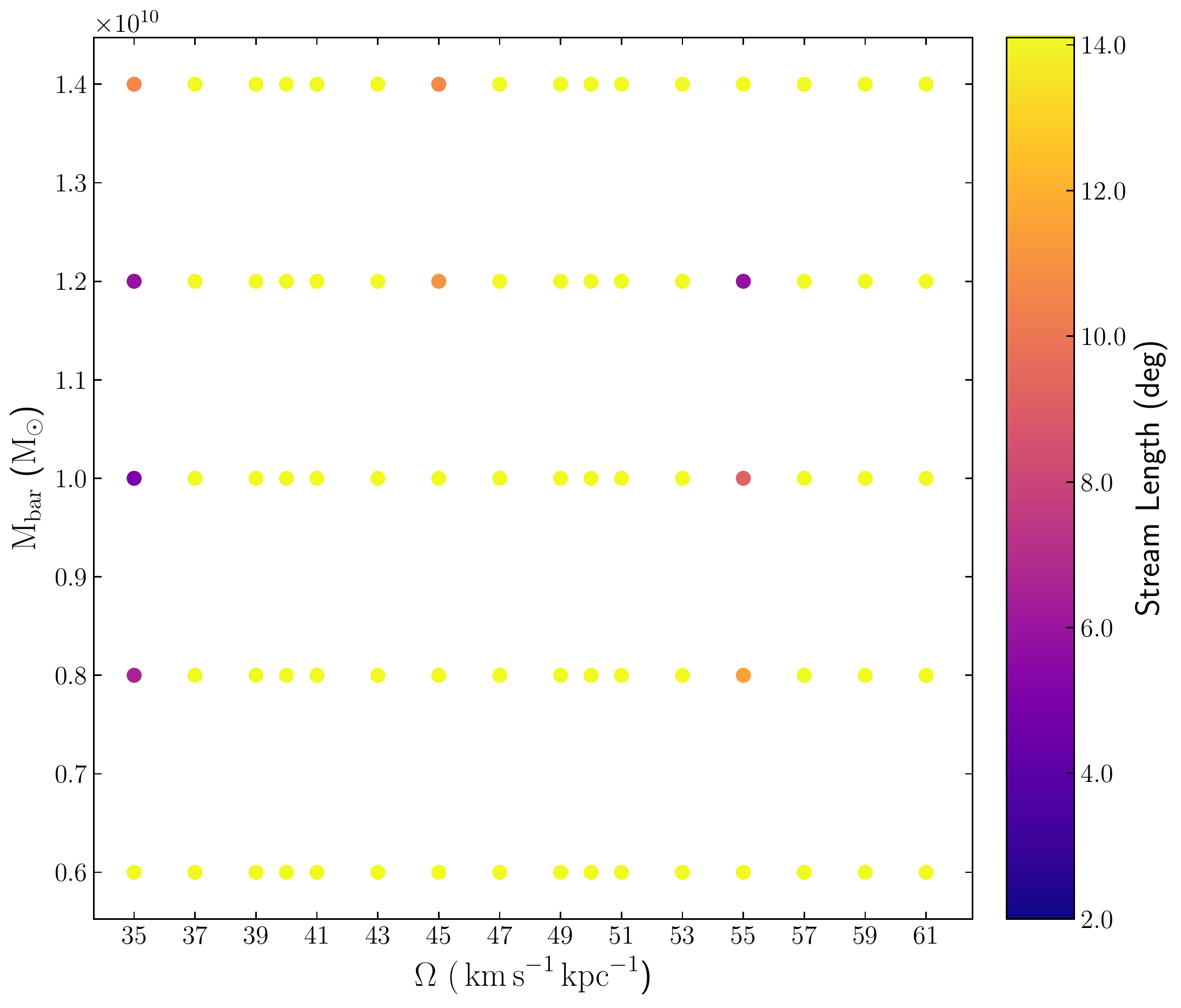}
\caption{Length of the Pal 5 stream as a result of varying the mass and pattern speed of the bar. For certain pattern speeds such as 35, 45 and 55 $\rm{km \ s^{-1} \ kpc ^{-1}}$, the stream is cut short if the mass of the bar is greater than $6\times 10^{9} \ \rm{M}_{\odot}$. For the case $\rm{M_{bar}} = 1.4\times 10^{10} \rm{M}_{\odot}$ and with a pattern speed of 55 $\rm{km \ s^{-1} \ kpc ^{-1}}$, the stream has a length of $\sim 14^{\circ}$ which is out of trend. This happened because a large number of stars from the leading arm are perturbed past the progenitor and they end up at the location of the trailing arm.}
\label{fig:stream_length}
\end{figure}

For each case of pattern speed and mass of the bar that we consider, we compute the stream density and its power. For certain pattern speeds, the Pal 5 stream is so heavily perturbed as to appear far shorter than what is observed, because many stars get large perturbations due to repeated interactions with the bar that remove them far from the observed portion of the stream. Therefore, we first compute the length of the stream and only consider bar models that do not lead to a significantly shorter stream. We define the length of the stream as the $\xi$ at which the stellar density drops below 20\% of the mean stellar density within $0.65^{\circ} < \xi < 3^{\circ}$ of the stream generated in the axisymmetric potential. The length of the Pal 5 stream for different pattern speeds and different bar masses is shown in Figure \ref{fig:stream_length}; the bar is assumed to be 5 Gyr old. We find that for a few combinations of pattern speed and bar mass the stream length is much shorter than what is observed. This effect of stream shortening could potentially be used to constrain the pattern speed of the bar, as pattern speeds that severely shorten the stream are disfavored. However, this may be degenerate with other parameters such as the dynamical age of the stream and this therefore requires a deeper investigation to become a useful constraint. For the remaining analysis, we remove the few pattern speeds for which the stream is severely shortened and only consider the cases which lead to a stream length that is comparable to the observed angular extent of the Pal 5 stream which is $\sim 14^{\circ}$ in $\xi$ \citep{Ibata2015}.

\begin{figure}
\includegraphics[width=0.5\textwidth]{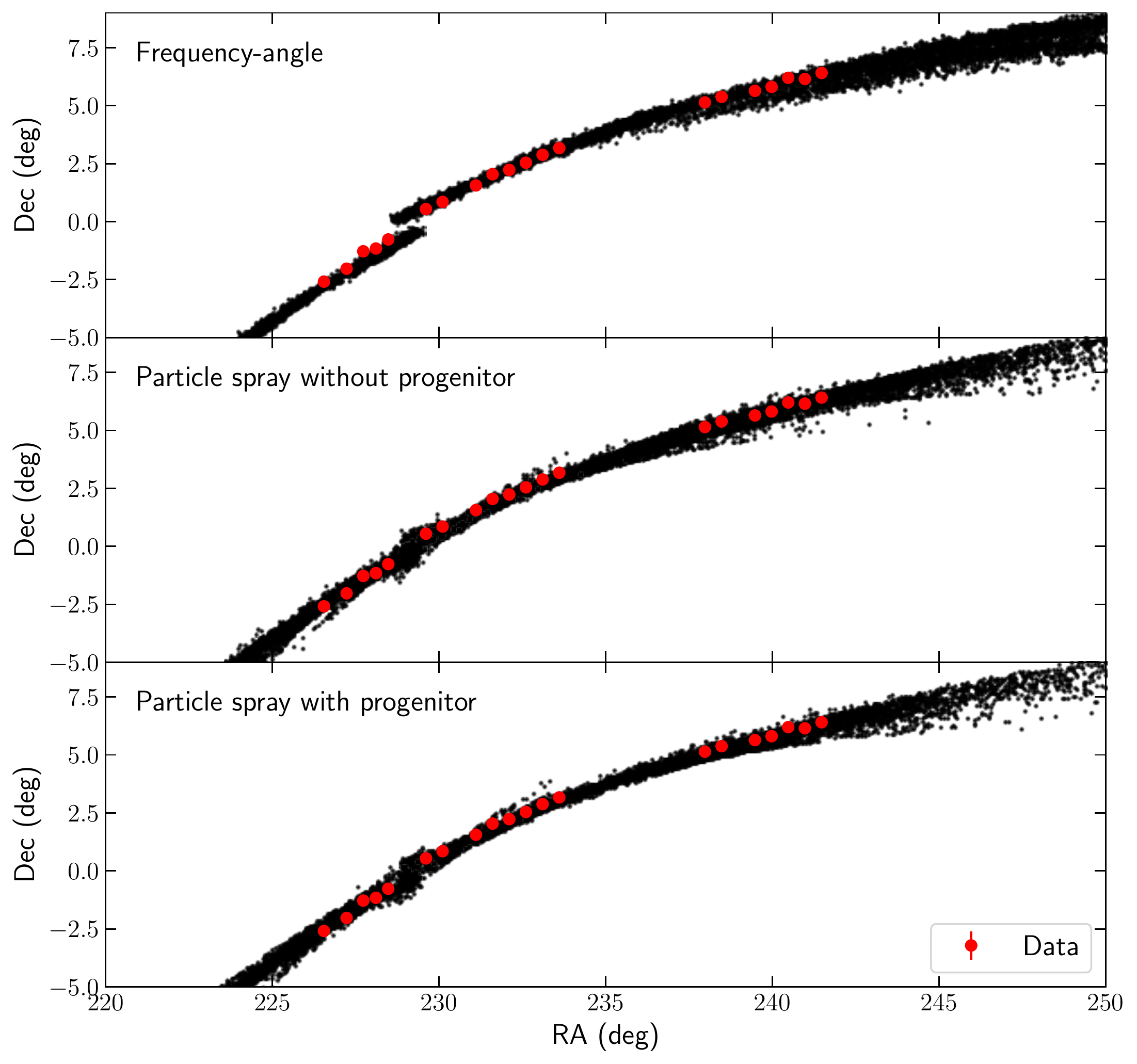}
\caption{Comparison of the present day location of the mock Pal 5 stream generated by the frequency-angle and \textit{particle spray} methods in a Milky Way potential with a bar of mass $10^{10}\ \rm{M}_{\odot}$ and pattern speed of 39 $\rm{km \ s^{-1} \ kpc ^{-1}}$. The overlaying red data points are from \citet{Fritz2015}. The top panel shows the stream in the frequency-angle framework following \citet{Bovy2014} which excludes the effects of the motion of the progenitor due to the bar. The middle and bottom panel shows the stream generated using the \textit{particle spray} technique without and with the bar's effect on the progenitor's orbit respectively.} 
\label{fig:streamdf_spraydf}
\end{figure}

\begin{figure}
\includegraphics[width=0.5\textwidth]{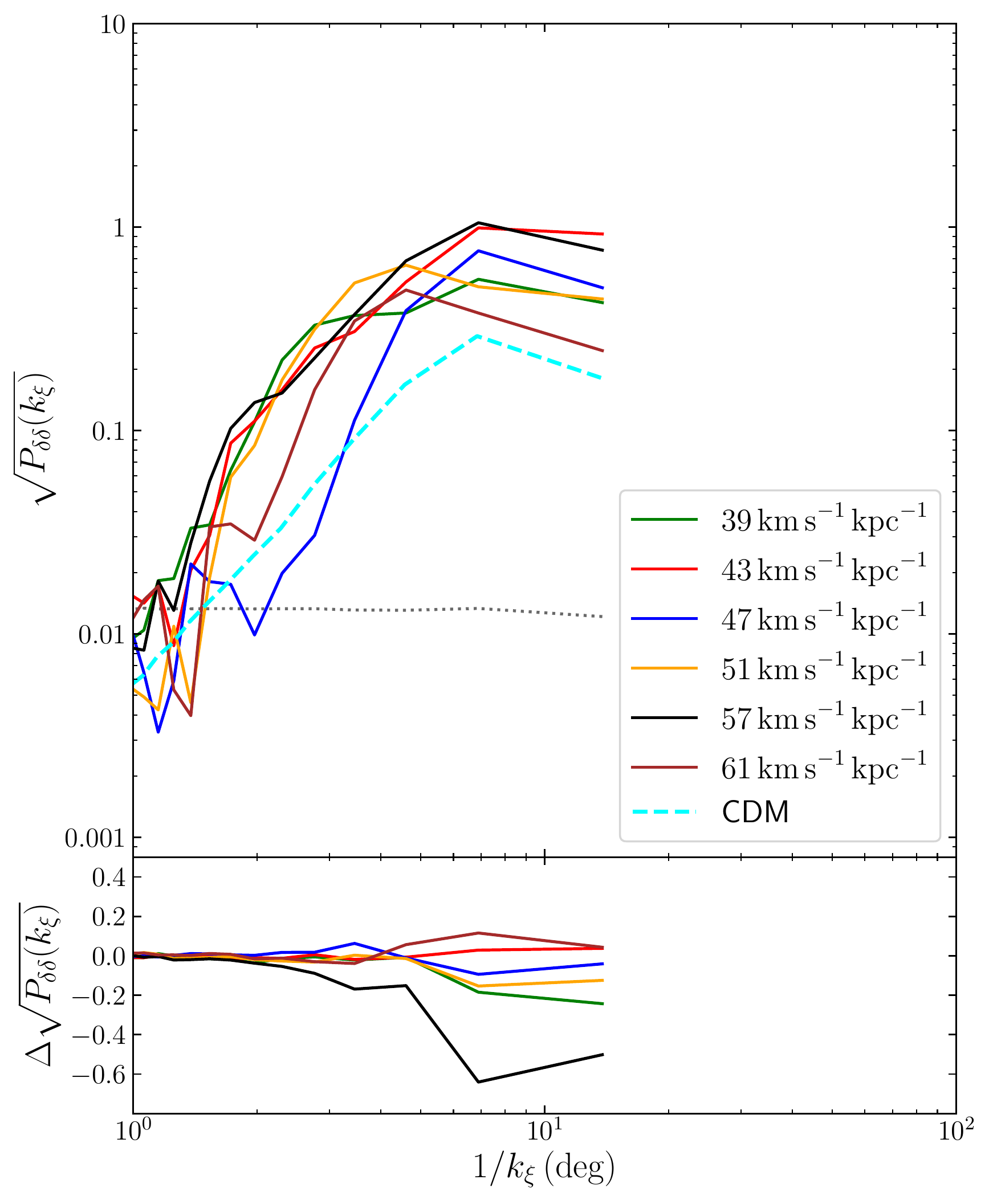}
\caption{Power spectrum of the density of the Pal 5 stream evolved in a barred Milky Way potential with different pattern speeds. We only consider pattern speeds for which the stream length is close to the observed length of the Pal 5 stream. In each case the bar is 5 Gyr old and has a mass of $10^{10} \rm{M}_{\odot}$. The top panel shows the power spectrum of the stream density. The gray dotted horizontal line shows the noise power as a result of the shot noise. The cyan dashed curve is the median power spectrum of 1,000 simulations of the stream density as result of impacts with CDM subhalos of mass in the range $10^{5} - 10^{9} \ \rm{M}_{\odot}$ from \citet{Bovy2016a} for comparison. The bottom panel displays the difference between the power in the case where the effect of the perturbation on the progenitor orbit is considered and the case in which it is not considered, as described in the text. The bar induces power on large scales that is similar or larger than that induced by dark matter subhalos, but drops significantly on small scales.}
\label{fig:power_patspeed}
\end{figure}

\begin{figure}
\includegraphics[width=0.5\textwidth]{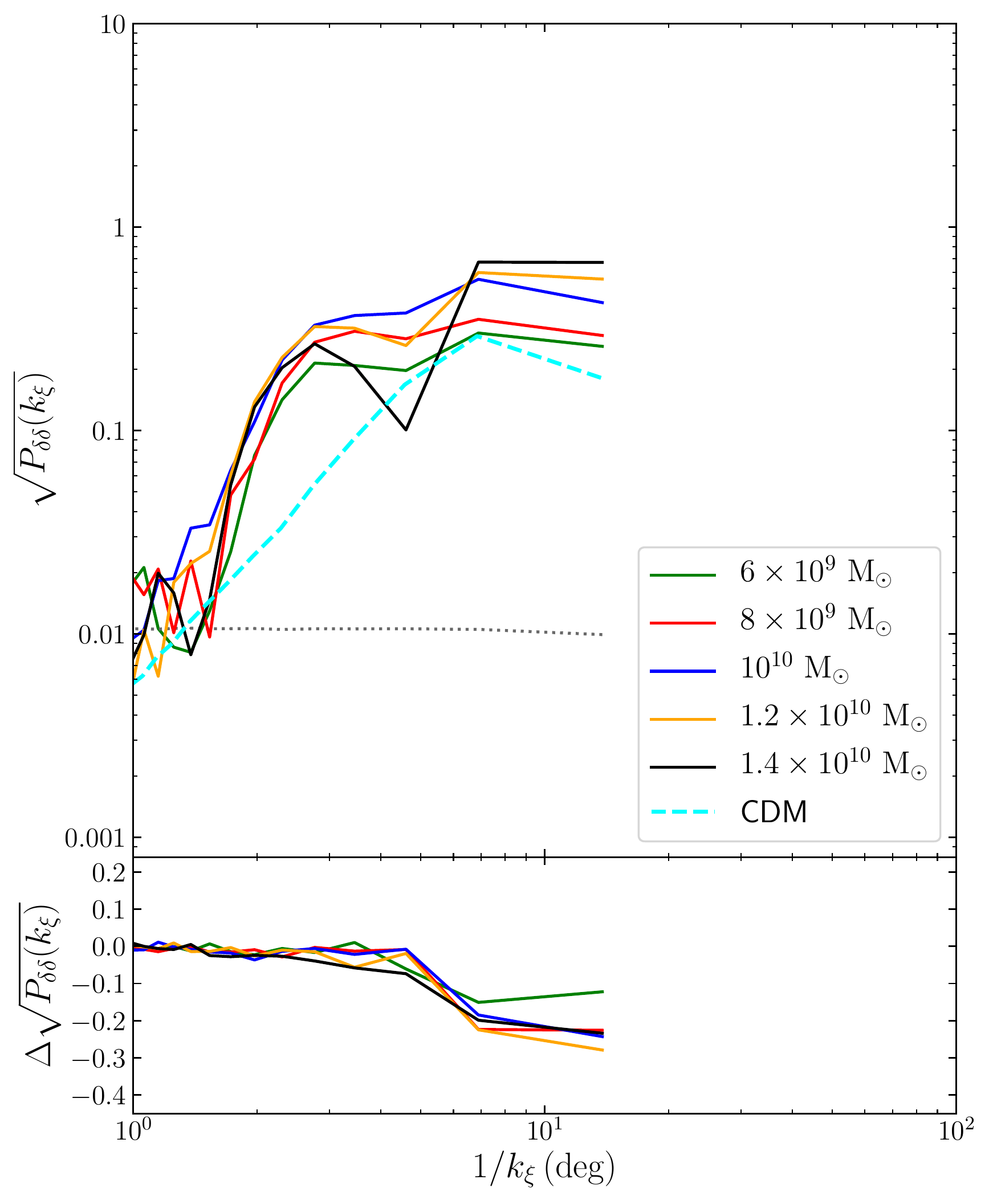}
\caption{Same as Figure \ref{fig:power_patspeed}, but showing the effect of varying the mass of the bar on the density power spectrum. The pattern speed of the bar is 39 $\rm{km \ s^{-1} kpc^{-1}}$.}
\label{fig:power_mass}
\end{figure}

\begin{figure}
\includegraphics[width=0.5\textwidth]{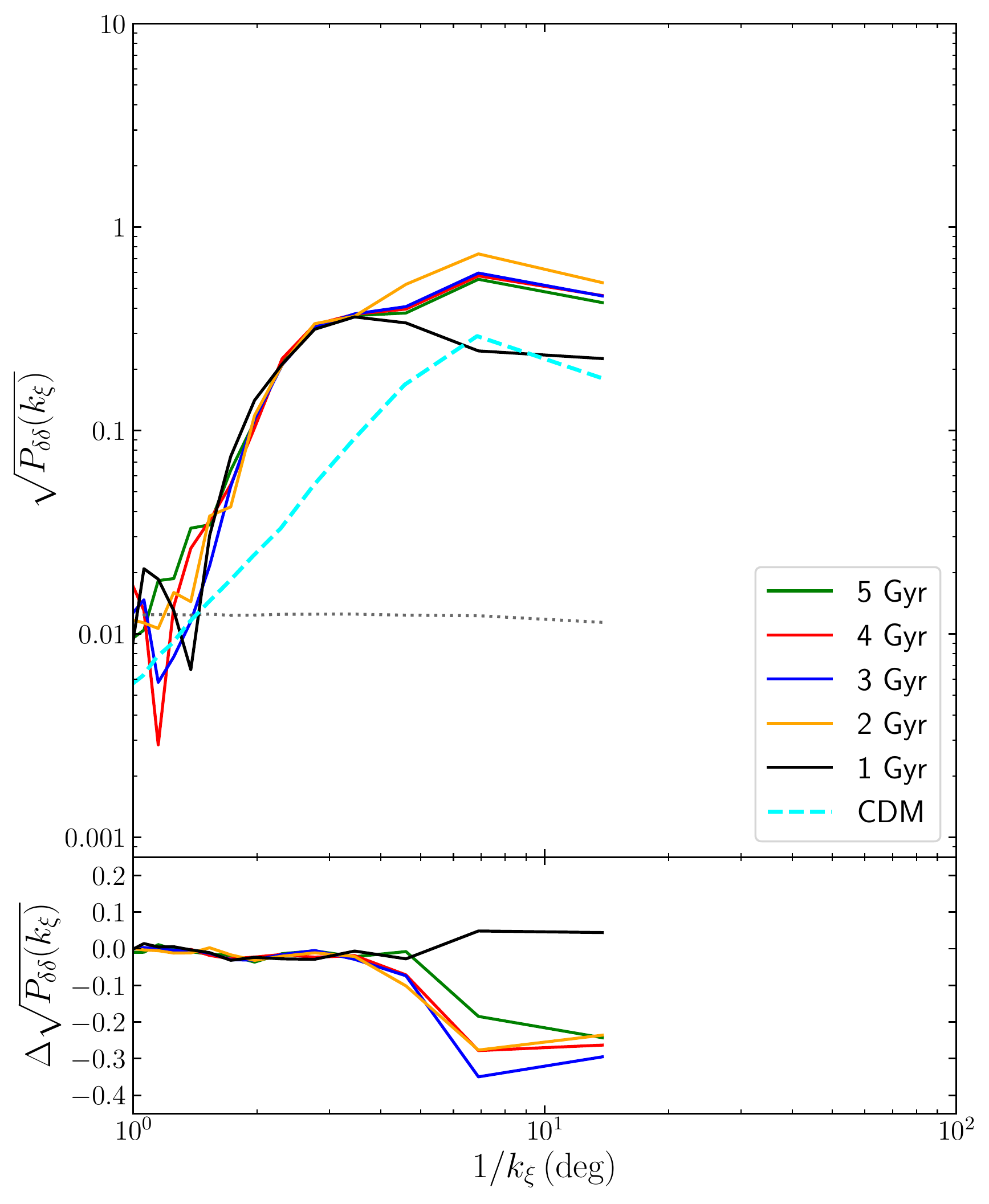}
\caption{Same as Figure \ref{fig:power_patspeed}, but showing the effect of varying the age of the bar on the density power spectrum. The mass of the bar is $10^{10} \, \rm{M}_{\odot}$ and its pattern speed is 39 $\rm{km \ s^{-1} kpc^{-1}}$. The age of the bar has only a minor effect on the induced power, especially on small scales.}
\label{fig:power_age}
\end{figure}

The approach for computing the density and its power spectrum described above does not take into account the fact that the progenitor's orbit is slightly different in the barred potential compared to that in the axisymmetric potential as shown in Figure \ref{fig:Pal5_orbit}. Therefore, the phase space coordinates of the stripped stars at the time of stripping will be different in the barred potential. This may lead to a different stellar density today along the stream. To investigate the effects of the progenitor's different orbit on the stream density, we generate mock streams using an implementation of the \textit{particle spray} technique described in \citet{Fardal2014}. In this approach, the Pal 5 progenitor is integrated back for 5 Gyrs from the present time in the barred Milky Way potential. Stripped stars are generated along the progenitor's orbit by offsetting them at the time of stripping from the progenitor in the instantaneous orbital plane (perpendicular to the angular momentum), with offsets in position and velocity some fraction of the tidal radius and circular velocity. These stripped stars are then integrated forward until the present time in the same barred potential. For consistency with the method with which the effect from dark matter subhalos and GMCs on the Pal 5 stream is computed, which uses the frequency-angle framework for mock stream generation (see below), we use the particle-spray mock streams only for determining the size of the effect of including the progenitor's perturbation on the resulting power spectrum---nevertheless, the induced power in the mock perturbed streams is similar with both methods.

Thus, to determine the impact of the perturbation to the progenitor's orbit, we compare the power spectrum of the above mock Pal 5 stream with that of a stream generated in the \textit{particle spray} technique but neglecting the deviations of the progenitor's orbit due to the barred potential. The latter is generated by stripping stars along the progenitor's orbit in an axisymmetric potential and then integrating them forward to the present time in the barred potential. Figure \ref{fig:streamdf_spraydf} compares the resulting stream generated by the different methods and demonstrates that the final stream looks similar in both frequency-angle and particle spray method. We quantify the effect of the progenitor's motion due to the bar by computing the difference in the power spectrum between these two cases $\Delta \sqrt{P_{\delta\delta}(k_{\xi})} = \sqrt{P_{\delta\delta}}|_{\rm{with \ progenitor}} - \sqrt{P_{\delta\delta}}|_{\rm{without \ progenitor}} $.

\subsection{Results}\label{sec:bar_result}

In Figure \ref{fig:power_patspeed}, we show the power spectrum for a 5 Gyr old bar of mass $10^{10} \, \rm{M}_{\odot}$, for different values of the pattern speed: 39, 43, 47, 51, 57, 61 $\rm{km \ s^{-1} \ kpc ^{-1}}$. The power at the higher end of the angular scale is very sensitive to the pattern speed of the bar. There is no clear trend of increase or decrease of power with pattern speed. This suggests that a resonance-like condition is responsible for the structure we see. This is similar to what is seen in simulations of the bar's effect on the evolution of the Ophiuchus stream \citep{Hattori2015} where the stream members in resonance with the bar suffer maximum torque from it, which results in more density perturbations. For the faster rotating bars with pattern speeds $\gtrsim 50\,\rm{km \ s^{-1} \ kpc^{-1}}$, the power is smaller compared to the other cases. This is interesting, because until recently such fast pattern speeds were the preferred value, because they explain the presence of the Hercules stream in the solar neighborhood \citep[e.g.,][]{Dehnen2000,Bovy2010,Hunt2017}. 

For the fiducial pattern speed of 39 $\rm{km \ s^{-1} \ kpc^{-1}}$ (green curve) the power is comparable to the power induced by dark matter subhalos, which is of the same order as the observed power of Pal 5. The predicted power from dark matter subhalos is shown by the thick dashed cyan line, which represents the median power spectrum of the stream density as a result of impacts with CDM subhalos in mass range $10^{5} - 10^{9} \, \rm{M}_{\odot}$. These CDM subhalo impacts were carried out following the same procedure as in \citet{Bovy2016a}. However, unlike in Bovy et al. (2017), who normalized CDM-perturbed mock streams using their unperturbed density, we normalize the CDM-perturbed streams using the same type of polynomial fit as we use for other perturbers and for the data for consistency's sake; this causes a small difference in the power on large scales when comparing our CDM curves to those of \citet{Bovy2016a}. The shot noise power for our bar simulations in all cases is at the level of $10^{-2}$ and is shown by the gray dotted horizontal line. The shot noise is a limitation stemming from only using 500,000 stream particles; the true power induced by the bar on small scales is below this noise floor and therefore smaller than that from dark matter subhalos. The bottom panel in Figure \ref{fig:power_patspeed} displays the difference in power when the effect of the progenitor's different orbit is considered. For a pattern speed of 57 $\rm{km \ s^{-1} \ kpc^{-1}}$ this effect is most prominent indicating considerable departure from the orbit in the axisymmetric potential. In most cases the power difference is below zero, indicating that including the effect of Pal 5's orbit should result in lowering the power, and so the power presented in the top panel are an overestimate, although on the logarithmic scale of the upper panel this difference is small. 

Next, we explore the effect of varying the mass of the bar on the power spectrum of Pal 5. In Figure \ref{fig:power_mass}, we show the power spectrum for different bar masses with pattern speed set to 39 $\rm{km \ s^{-1} \ kpc ^{-1}}$. The sub-panel in each figure again displays the difference in the power spectrum $\Delta \sqrt{P_{\delta\delta}(k_{\xi})}$ due to the effect of the bar on the progenitor's orbit. There is a clear trend of increasing power with the mass of the bar. This is expected, because a bar with more mass imparts stronger perturbations to the stream. In this case, including the progenitor's motion lowers the power by almost the same amount for all the different mass bars.

The effect of varying the age of the bar on Pal 5 stream's power spectrum is shown in Figure \ref{fig:power_age}. It was shown in \cite{Cole2002}, that the Galactic bar is less than 6 Gyr old and likely less than 3 Gyr. We vary the age of the bar between 1 and 5 Gyr and compute the power spectrum of Pal 5 in each case. The power of Pal 5 is virtually unaffected by the age of the bar as long as it is at least 2 Gyr old.

\section{Effect of the spiral arms on Pal 5}
\label{sec:spiral}

\subsection{Spiral structure models}

In this section we investigate the possible effect of spiral structure on the density of Pal 5 stream. We model the gravitational potential due to spiral structure using \texttt{galpy}'s \texttt{SpiralArmsPotential} which is based on the analytic model of \cite{Cox2002}. The potential has the following form :

\begin{multline}
\Phi(R,\phi,z) = -4\pi G H \rho_{0} \exp\left(\frac{r_{0} - R}{R_{s}}\right)\\
\times \sum_{n}\left(\frac{C_{n}}{K_{n}D_{n}}\right) \cos(n\gamma) \left[\sech\left(\frac{K_{n}z}{\beta_{n}}\right) \right]^{\beta_{n}}
\end{multline}

where 

\begin{equation}
K_{n} = \frac{nN}{R\sin(\alpha)} 
\end{equation}
\begin{equation}
\beta_{n} = K_{n}H(1+0.4K_{n}H)
\end{equation}
\begin{equation}
D_{n} = \frac{1 + K_{n}H + 0.3(K_{n}H)^{2}}{1 + 0.3K_{n}H}
\end{equation}

\begin{equation}
\gamma = N\left[\phi - \phi_{\rm{ref}} - \frac{\ln(R/r_{0})}{\tan(\alpha)}\right]
\end{equation}
N denotes the number of spiral arms, $\rho_{0}$ sets the amplitude, and $r_{0}$ is a reference radius, which we took to be 8 kpc. The pitch angle $\alpha$ is set to $9.9^{\circ}$ and the reference angle $\phi_{\rm{ref}}$ is set to $26^{\circ}$ \citep{Siebert2012,Faure2014,Monari2016a}. $R_{s}$ is the radial scale length of the spiral density which we set to 3 kpc, similar to \texttt{MWPotential}'s (effective) exponential disk scale length which is $\approx2.6$ kpc \citep[][Table 1]{Bovy2015}
and $H$ is the vertical scale height set to 0.3 kpc. The $C_{n}$ determine the profile of the spiral arms: if $C_{n} = 1$, then we get a sinusoidal potential profile, whereas if $C_{n} = [8/3\pi, 1/2, 8/15\pi]$ then the density takes approximately a cosine squared profile in the arms with flat interarm separations. Following \citet{Monari2016}, the amplitude $\rho_{0}$ is set such that the radial force at the location of the Sun due to the spiral arms is around one percent of the radial force due to the axisymmetric Milky Way potential (\texttt{MWPotential2014}). We explore the effects of varying the following parameters: (a) number of arms, either $N=2$ or 4 , (b) amplitude such that the local radial force is $0.5 \%$ or $1 \%$ of the total local radial force \citep{Monari2016}, and (c) pattern speed of 19.5 and 24.5 $\rm{km \ s^{-1} kpc^{-1}}$ on Pal 5's density power spectrum.

\begin{figure}
\includegraphics[width=0.5\textwidth]{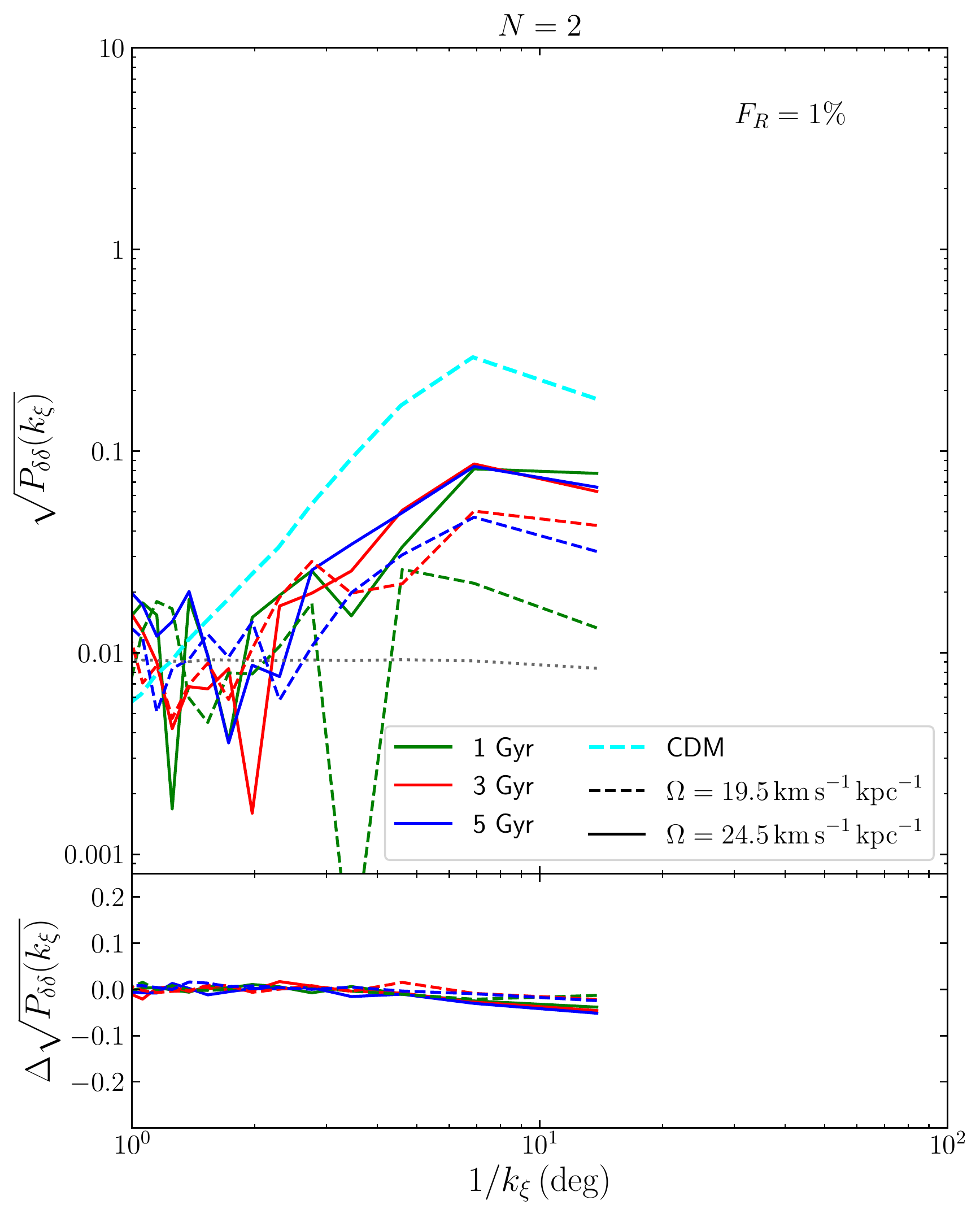}
\caption{Same as Figure \ref{fig:power_patspeed} but for a Milky Way potential that includes spiral structure rather than a bar. The plot shows the effect of varying the age and pattern speed of a two-armed spiral that contributes $1\,\%$ of the radial force at the Sun.}
\label{fig:spiralN2}
\end{figure}

\begin{figure}
\includegraphics[width=0.5\textwidth]{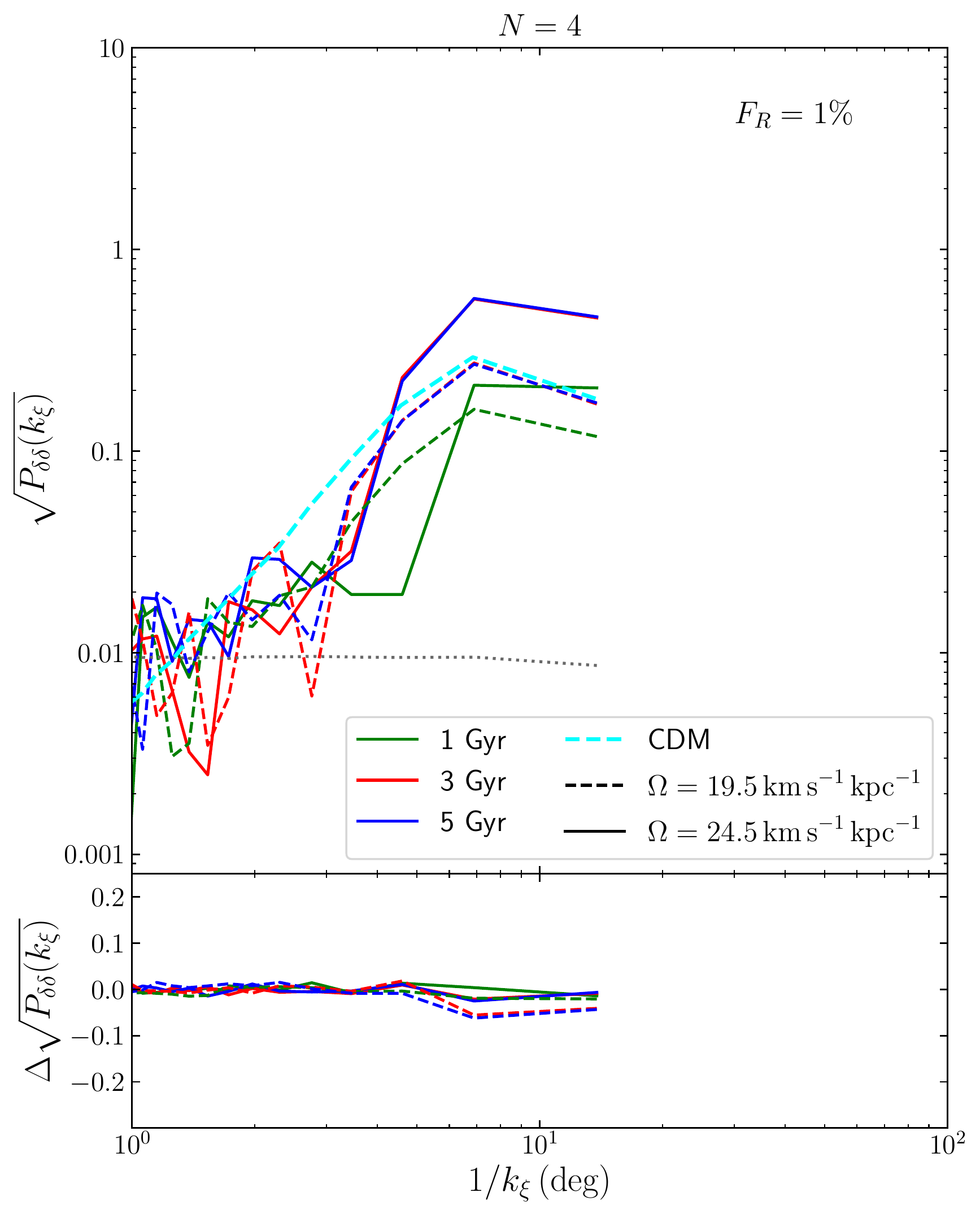}
\caption{Same as Figure \ref{fig:spiralN2} but for a four-armed spiral potential.}
\label{fig:spiralN4}
\end{figure}

As for the bar above, we  grow the amplitude of the spiral potential from zero to full over two rotation periods of the spiral arms following the prescription of \citet{Dehnen2000}. The spiral potential is in all cases added to the axisymmetric \texttt{MWPotential2014}. We then follow the same set of steps as for the Galactic bar above. The results are shown in Figures \ref{fig:spiralN2} and \ref{fig:spiralN4} for 2 arms and 4 arms spiral potential respectively. In each case the amplitude is set such the local radial force from the spirals is 1\% of the radial force of the background axisymmetric potential. For the lower value of the radial force, we found the power to be consistently lower than all the 1\% cases, as expected, and hence we do not show them.

\begin{figure}
\includegraphics[width=0.5\textwidth]{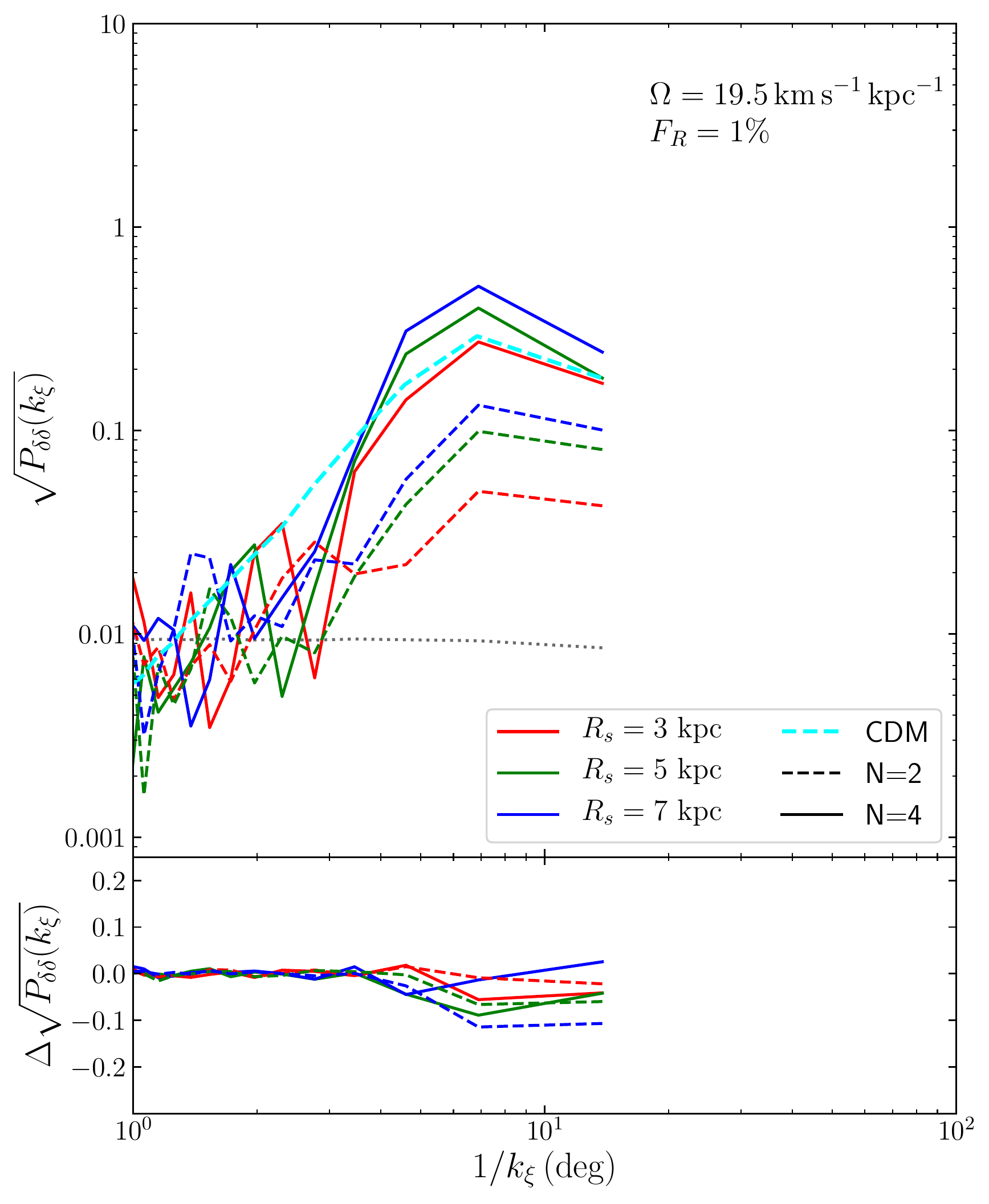}
\caption{Same as Figure \ref{fig:spiralN2} but comparing the power induced when the scale length of the spiral pattern, $R_{s}$, is varied. For all cases the pattern speed is set to $19.5\,\mathrm{km\,s}^{-1}\,\mathrm{kpc}^{-1}$ and the amplitude is set such the radial force from the spirals is 1\% of the radial force of the background axisymmetric potential at the Solar radius. There is a clear trend of more density power as the scale length is increased. Only if the spiral scale length is larger than the disk scale length do spirals significantly affect Pal 5's density. }
\label{fig:spiral_varyRs}
\end{figure}

\subsection{Results}\label{sec:spiral_effect}

From Figure \ref{fig:spiralN2}, it is clear that for a two-armed spiral arm potential contributing $1\%$ of the radial force at the Sun, the power induced on the Pal 5 stream is around 3 times lower than the power induced due to CDM subhalo impacts at large scales. Varying the age and the pattern speed does not show any strict trend in the power. However, varying the age of the spiral arms above 3 Gyr has almost negligible effect on the power. The power difference in the subplot implies that the motion of the Pal 5 progenitor has very little effect on the progenitor's orbit and leads to lowering of the power. Increasing the number of spiral arms to four significantly increases the power induced in the Pal 5 stream as shown in Figure \ref{fig:spiralN4}. A 3 or 5 Gyr old spiral arms results in identical power at the large scales regardless of the pattern speed. In this case, the motion of the progenitor again has almost no effect on the power. At large scales, the power induced due to the CDM subhalo impacts is consistent with that due a four-armed spiral arm that is at least 3 Gyr old with a pattern speed of $19.5\,\mathrm{km\,s}^{-1}\,\mathrm{kpc}^{-1}$. Figure \ref{fig:spiral_varyRs} compares the power induced when the exponential scale length, $R_{s}$, of the spiral arms is varied which shows a clear trend of increasing power with scale length. This is expected as the density of the spiral arms remains high at larger Galactocentric radial distances for longer scale lengths. Only if the spiral scale length is large do spirals have a large effect on Pal 5's density.  

\begin{figure}
\centering
\includegraphics[width=0.5\textwidth]{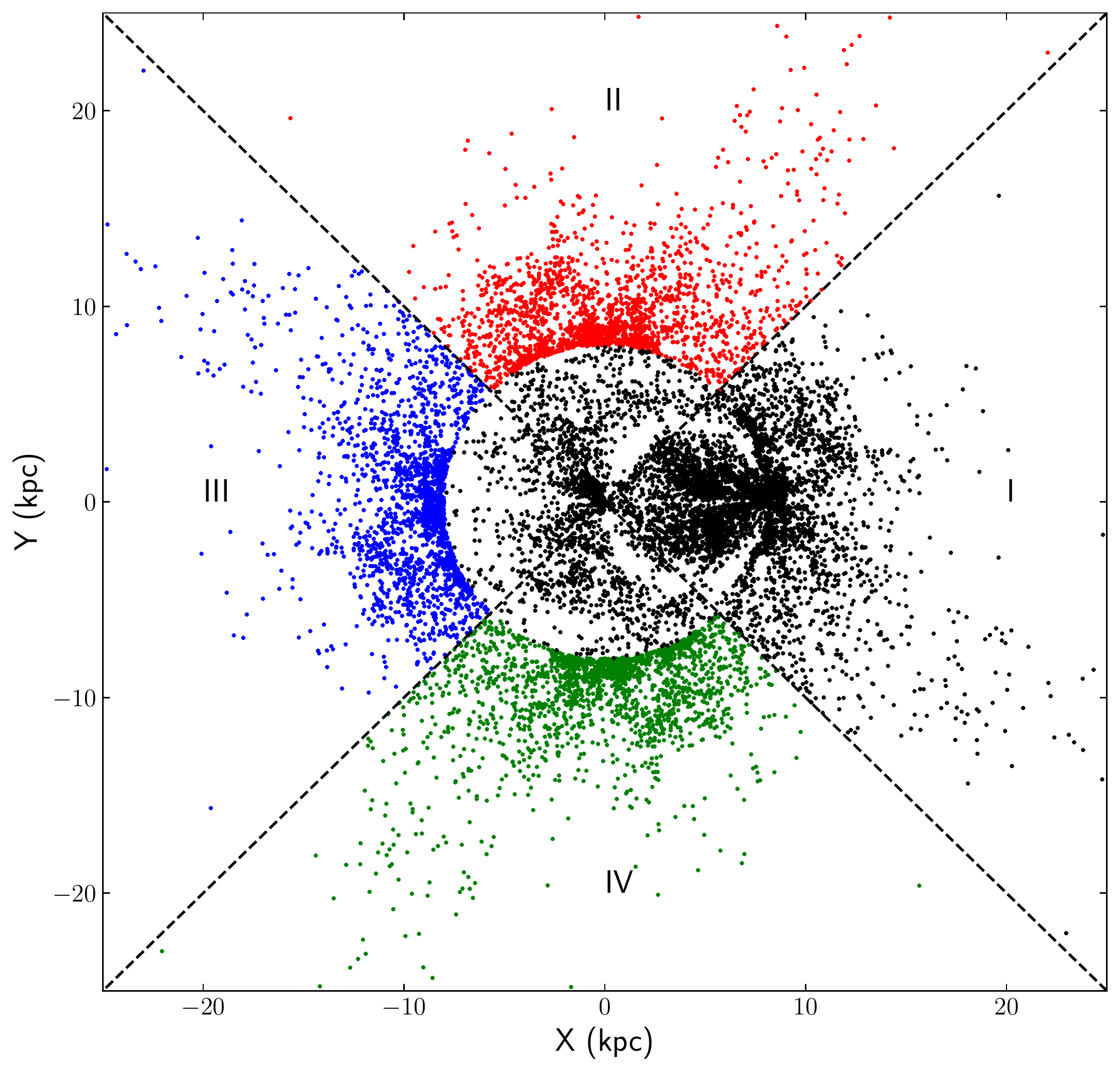}
\caption{Map of the GMCs after the empty patches in the outer disk have been filled by removing the outer disk GMCs from quadrants ``II", ``III", and ``IV" and  replacing them by the outer disk GMCs from quadrant ``I". The figure shows only the GMCs with Galactocentric radius less than 20 kpc since the apogalacticon of the Pal 5 stream is $\sim 14$ kpc. }
\label{fig:GMC_map}
\end{figure}

\section{Effect of the giant molecular clouds}
\label{sec:GMC}

\subsection{Modeling the Milky Way's population of GMCs}

In this section, we explore how the Galactic population of giant molecular clouds (GMCs) affects the Pal 5 stream. \citet{Amorisco2016} demonstrated that GMCs confined to the razor thin disk can impact globular cluster streams such as Pal 5 and give rise to gaps in their density. Because the size and mass of the largest GMCs is similar to that of low-mass dark matter subhalos, GMC-induced gaps are similar to the ones that result from dark matter subhalo impacts and therefore will introduce large uncertainties when using stellar streams as probes for dark matter subhalos. We investigate the cumulative effect of gravitational encounters of GMCs with the Pal 5 stream over its dynamical age by computing the power spectrum of density perturbations induced by GMCs rather than dark matter.

Rather than using a simple model of the GMC population in the Milky Way, we directly use a recent catalog of 8,107 GMCs from \citet{Miville-Deschenes2016}, which is close to complete for the largest GMCs that are of highest interest here (as we discuss below, GMCs with masses $\lesssim 10^5\rm{M}_\odot$ have very little effect). The map of the GMCs show several patches in the outer disk (Galactocentric radius > $R_0$) on the other side of the Galactic center that are devoid of GMCs which is due to the difficulties in observing GMCs on the other side of the Galactic center. To better localize these empty patches, we divide the GMC map into four quadrants centered at the Galactic center. The first quadrant contains the Sun and has a fairly uniform distribution of GMCs in the outer disk. To fill the empty patches in the other three quadrants, we remove the outer disk GMCs and copy the outer disk GMCs from the first quadrant into them. Figure \ref{fig:GMC_map} shows the map of the GMCs after the empty patches have been filled which increased the total number of GMCs to 14,542 within a Galactocentric radius of 40 kpc. The colored points indicate the GMCs that were copied from the first quadrant. We do not correct the GMC catalog within the solar radius, because much of the non-uniform spatial distribution there is likely mostly due to non-axisymmetric motions of the GMCs affecting their inferred distance rather than incompleteness. We setup their orbits by positioning them at their present-day location in the Galaxy and placing them on a circular orbit in the \texttt{MWPotential2014} potential. We then evolve them back in time for the dynamical age of the Pal 5 stream. Next, we evolve both the GMCs and the Pal 5 stream forward in the same potential and compute the impacts of the GMCs on the stream during this time. 

The mass $M$ and the physical radius $R$ of each GMC in the \citet{Miville-Deschenes2016} catalog corresponds to its entire angular extent on the sky. However, the radius $R$ is not the scale radius, but the full radius, and therefore we model the GMCs as Plummer spheres with scale radius equal to one third the full radius because for a Plummer sphere, 90\% of mass is contained within 3 times the scale radius.  

\begin{figure}
\includegraphics[width=0.5\textwidth]{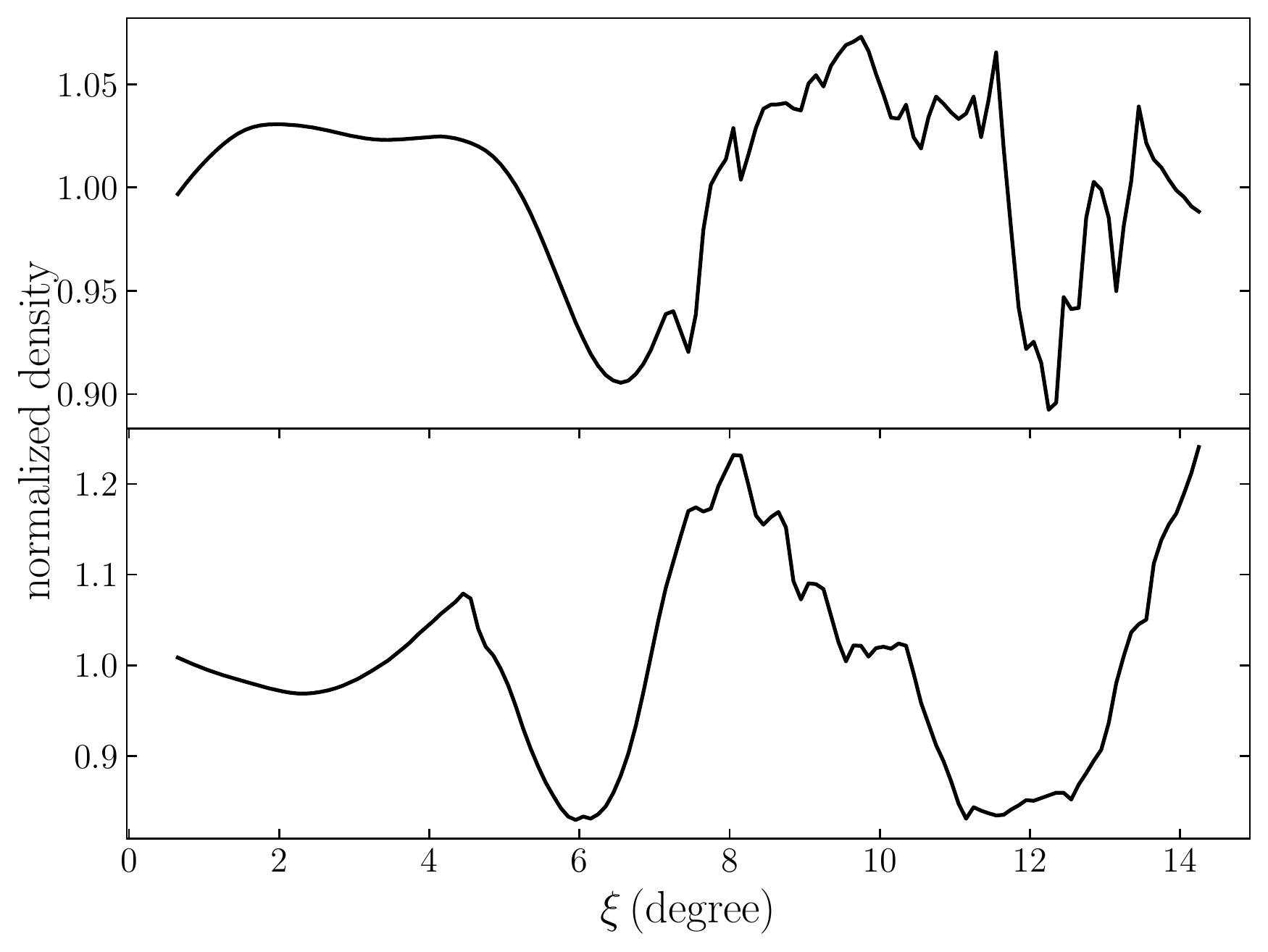}
\caption{Normalized stream density of two realizations from the 40 different realizations of the Milky Way GMC population as explained in the text. There are a number of small scale perturbations which results in high power at small angular scales as shown in Figure \ref{fig:GMC_powerdisp}.}
\label{fig:GMC_dens}
\end{figure}

To compute the effect of the GMCs on Pal 5 stream, we only consider GMCs with $M > 10^{5} ~ \rm{M}_{\odot}$, because we found that including the lower mass GMCs resulted in negligible change in the power. Following \citet{Bovy2016a}, we consider impacts up to a maximum impact parameter $b_{\rm{max}} = 5\times r_{s}(M)$. This takes into account the effect that smaller (low mass) GMCs need to pass more closely by the stream compared to bigger (more massive) ones to have an observable effect. The GMC impacts are modeled by the impulse  approximation and the resulting stream density is computed using the line-of-parallel-angles approach as described in \citet{Bovy2016a}. Following the same reference, to save computational time, we re-sample impacts on a discrete grid of time over the dynamical age of the stream. To properly resolve the interactions between the GMCs and the stream, it is necessary to compute the impact parameters---time of impact, closest approach---with a time resolution at least equal to the typical time scale over which a GMC interacts with a stream, which is of order $r_{s}/v \simeq (\rm{few} \ 100 \rm{pc})/(200 \ \rm{km/s}) \sim 1 \ Myr)$. We have checked that the density power has converged using this time resolution for computing impact parameters. 

The lifetime of a typical GMC is between 10 to 50 Myr \citep{Jeffreson2018}, much longer than the typical duration of their interaction with a stream and therefore we are justified in treating them as having a fixed mass. This also means that the present day population of GMCs did not exist during the entire dynamical lifetime of the Pal 5 stream and is thus at best a proxy for the population of GMCs that may have interacted with the stream. To compute the effect of an evolving population of GMCs impacting the stream, we create new realizations of the GMC population by adding random rotations to the Galactocentric cylindrical $\phi$ coordinates of the present GMCs and then follow the same steps as above to find the overall density perturbations imparted on the stream. This $\phi$ randomization maintains the spatial and mass distribution of GMCs, but allows us to study the range of possible histories of GMC interactions. We generate 40 different random $\phi$ realizations. Figure \ref{fig:GMC_dens} displays the resulting density contrast in two cases. 

\begin{figure}
\includegraphics[width=0.5\textwidth]{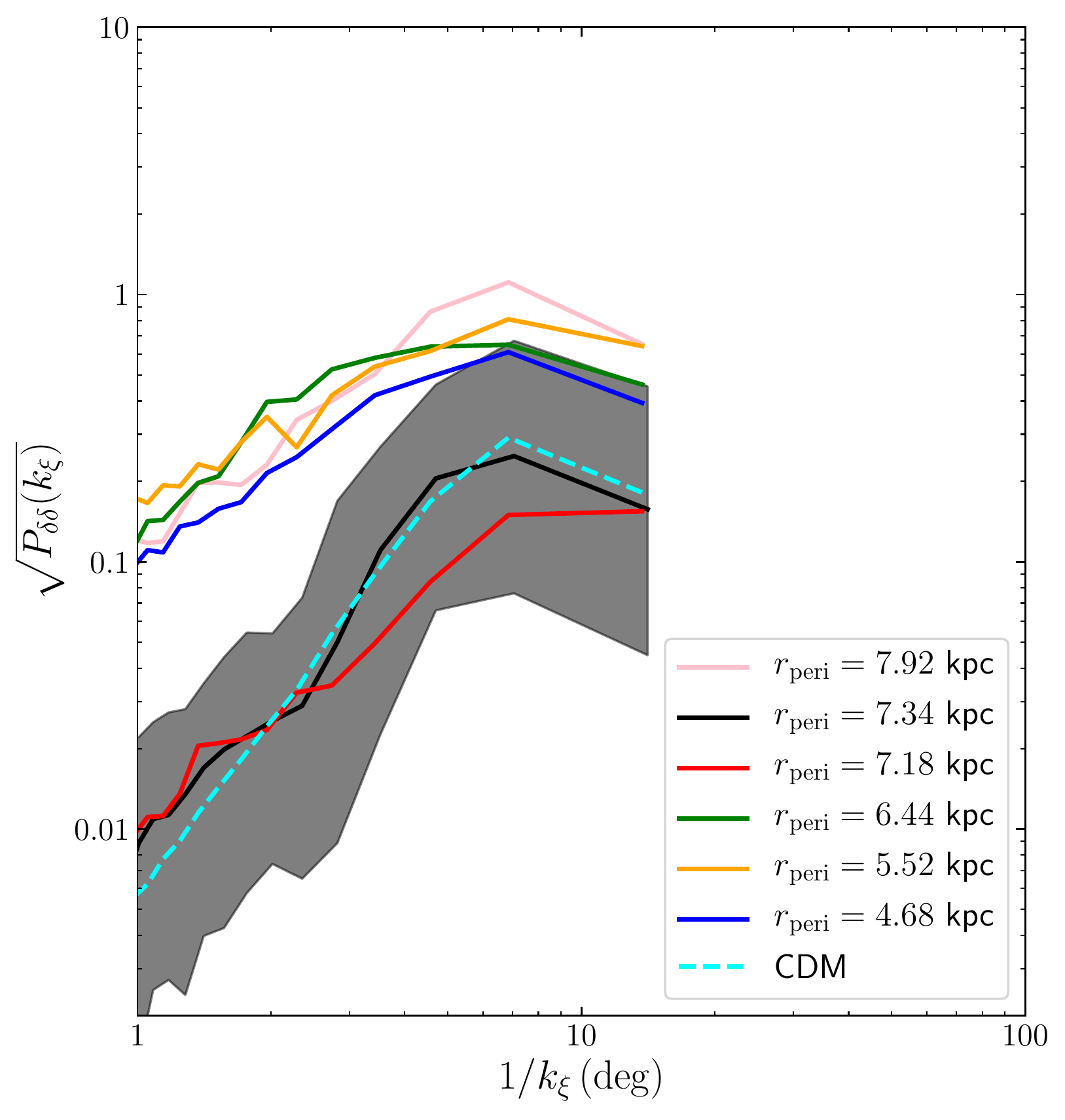}
\caption{Median power spectrum of density perturbations from GMCs for the fiducial Pal 5 orbit (pericenter = 7.34 kpc) from 42 different realizations of the Galactic population of GMCs with mass greater than $10^{5} \ \rm{M}_{\odot}$ (black curve); the shaded region displays the $2\,\sigma$ range spanned by the 40 realizations. Each colored curve represents the median power (over 40 realizations) of the Pal 5 stream with orbits with different pericenter radii. The cyan dashed curve shows the power of the stream density as a result of CDM subhalo encounters. Unlike the simulations that we performed for the bar and spiral arms, the density computed using the frequency-angle method does not contain numerical noise, so the noise curve that was present in the previous power spectra in this paper is absent here.}
\label{fig:GMC_powerdisp}
\end{figure}

\begin{figure}
\includegraphics[width=0.5\textwidth]{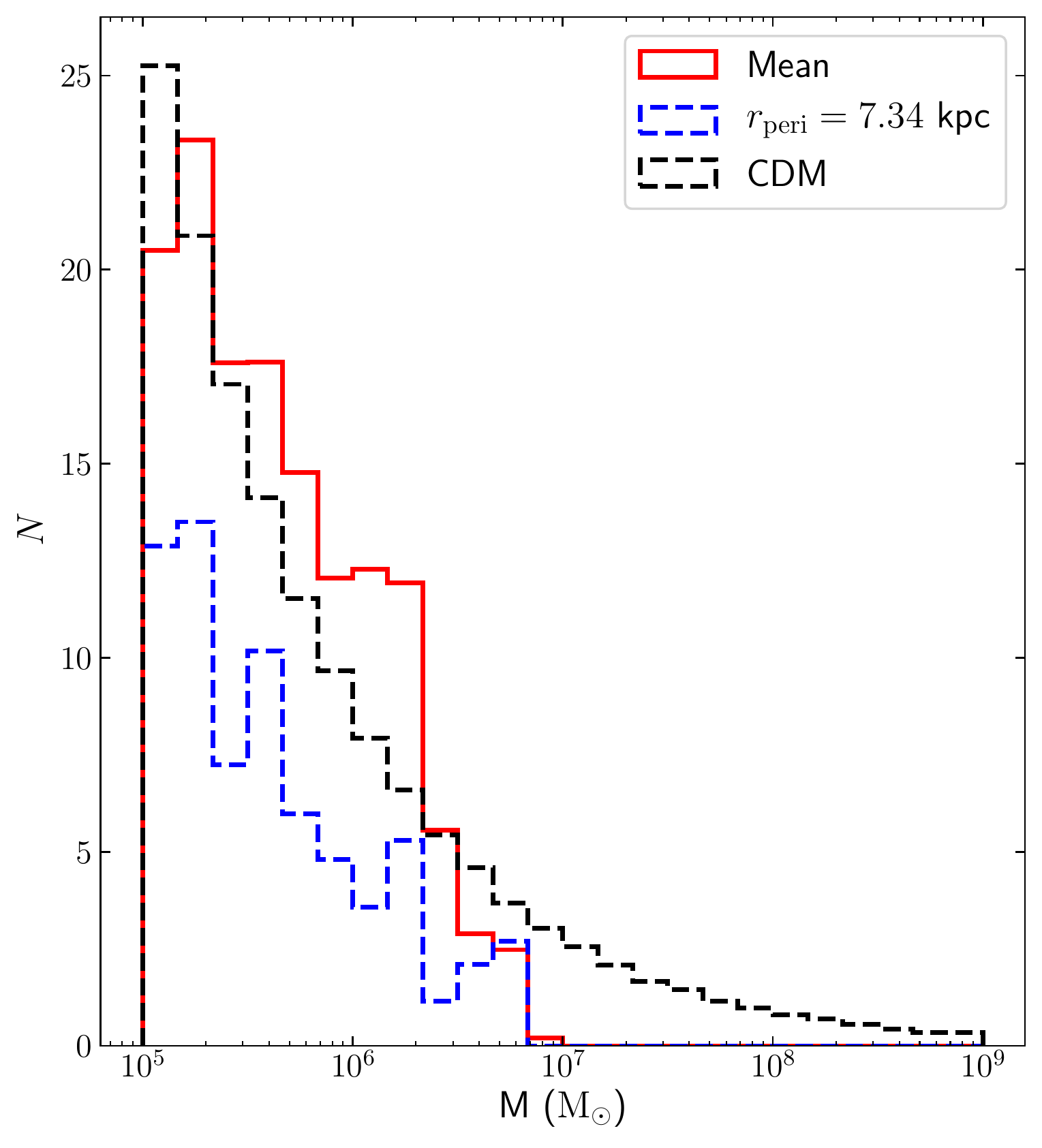}
\caption{Histogram showing the average number of impacts for different mass GMCs/subhalos that the Pal 5 stream encounters in different setups. The red solid line denotes the mean number of impacts over the 5 non-fiducial Pal 5 orbits (each orbit had 21 realizations) whose median powers are shown in Figure \ref{fig:GMC_powerdisp}. The blue dashed line denotes the mean number of GMC impacts of the fiducial Pal 5 orbit over the 40 realizations. The black dashed line denotes the mean number of impacts over the 1000 CDM-like simulations whose median power is shown by the dashed cyan curve in Figure \ref{fig:GMC_powerdisp}.}
\label{fig:GMC_CDM_impact}
\end{figure}

\subsection{Effect of varying Pal 5's pericenter} 
\label{sec:Pal5_peri}

\citet{Bovy2016} performed a detailed investigation of the orbits and Milky Way potential models that are consistent with the Pal 5 stream and other dynamical data in the Milky Way. The full range of possible Pal 5 progenitor's phase-space coordinates were sampled using MCMC. We use all of the generated MCMC chains to explore differences in Pal 5's orbit from our fiducial orbit model. We find that Pal 5's pericenter varies between 4.68 and 8.01 kpc; for the fiducial orbit that we have been considering so far, the pericenter radius is 7.34 kpc. Because GMCs are distributed non-uniformly in radius, with especially a much larger number of high-mass GMCs at radii $\lesssim 7$ kpc, variations in Pal 5's pericenter radius can have a big impact on the predicted effect from GMCs. Therefore, we consider 5 different pericenter values in the allowed range: 4.68, 5.52, 6.44, 7.18, and 7.92 kpc. To study these, we randomly pick 5 chains that correspond to these values out of all the MCMC chains from all the potentials. For each chain and its corresponding potential model, we follow the same procedure as for the fiducial orbit and potential, and compute GMC impacts  as described above. Finally, for each chain, we impact the stream in each case with 40 different realizations of the GMC population by adding random $\phi$ rotations to their current coordinates as described above. The resulting power spectrum for all the cases are shown in Figure \ref{fig:GMC_powerdisp} by the colored curves.   

\subsection{Results}\label{sec:GMC_results}

In Figure \ref{fig:GMC_powerdisp} we show the median power spectrum of the 40 GMC realizations and their $2-\sigma$ scatter (gray-shaded region) for the fiducial Pal 5 orbit. The median power at the largest scale is comparable to that of the CDM case. 
The lower bound is an order of magnitude less, indicating the wide range of power that the GMCs can impart to the stream. Compared to the CDM case, there is slightly more power due to the GMCs at lower angular scale. This seems counterintuitive in the light of Figure \ref{fig:GMC_CDM_impact}, which indicates the stream has a similar number of hits by low mass ($< 10^{6} \ \Msun$) subhalos (dashed black line) as by GMCs (dashed blue line). The difference arises because the GMCs are much more compact ($\sim 5$ times) than the subhalos, so they are capable of inflicting small scale perturbations to the stream, as also seen in their density in Figure \ref{fig:GMC_dens}. This difference in power at small scales can be used to statistically set the GMC impacts from CDM subhalo impacts. 

Varying the pericenter of the stream, we find that for orbits with pericenters that are less than the fiducial 7.34 kpc, the stream encounters many more GMC impacts. As a result the power in all these cases is much higher. In general, the number of GMCs increases as one goes closer to the Galactic center. However, the number of impacts depends not only on the number of GMCs, but also on their orbit relative to the stream, because that decides whether a GMC will fly by the stream with an impact parameter less than $b_{\rm{max}}$. 

The upper limit of the $2-\sigma$ dispersion of power of all the different pericenter cases is at the level of $\sim 1.2$, which is higher than the observed power of the Pal 5 stream. From all these results, we can conclude that our ignorance of the evolution of the GMCs over the dynamical age of the stream makes them the biggest source of uncertainty in using the Pal 5 stream as a probe for dark matter subhalos. 

\section{Impacts due to the globular clusters}
\label{sec:GCs}

The final baryonic component whose effect on the Pal 5 stream we consider is the population of globular clusters (GCs). The Milky Way hosts 157 GCs \citep{Harris1996,Harris2010}, which as dense, massive concentrations of stars may affect stellar streams. To determine their effect on the Pal 5 stream, we follow the same procedure as for the GMCs above. 

We obtain approximate orbits for the GCs as follows. For 75 of the GCs, we use the proper motions and other kinematic information from \citet{GaiaCollaboration2018}, who determined proper motions of these GCs using data from \textit{Gaia} Data Release 2 (DR2). For 72 of the remaining GCs, we obtain the same information from the recent catalog by \citet{Vasiliev2018}, who similarly used \emph{Gaia} DR2 data. For the remaining 10 GCs we were unable to find complete kinematic information and we do not consider them further. Aside from the proper motions, most of the phase-space coordinate information in both the \citet{GaiaCollaboration2018} and \citet{Vasiliev2018} comes from the \citet{Harris2010} globular cluster catalog, aside from some minor modifications from \citet{Baumgardt2018}. 

We obtained masses for 112 of the GCs in the sample from the catalog by \citet{Baumgardt2018}. For GCs without a mass measurement in this catalog, we conservatively assign them the highest of any GC in the catalog: $3.5\times 10^{6} \ \rm{M}_{\odot}$. Just like their kinematic information, the angular size of the GCs are taken from their respective catalogs and their physical radius then follows from multiplying the angular size by the distance.  We then model the GCs as Plummer spheres with scale radius set equal to their physical radius. As in the case of GMCs, by modeling the GCs as Plummer spheres, we make the assumption that most ($\sim 90\%$) of the mass is within their angular size and so the scale radius is set by dividing the angular radius by 3. Using the kinematic information, we compute the past orbit of each GC in \texttt{MWPotential2014} and use the same steps as for the GMCs to compute their impacts with the Pal 5 stream. Given that the maximum size of the GCs is $\sim 100$ pc, we consider impacts out to 0.5 kpc from the stream in all cases. 

The density perturbation arising from GC impacts on the Pal 5 stream are very small. The power of the relative density fluctuation is $\lesssim10^{-3}$ on all scales. This is below the contribution from all of the other baryonic perturbers. This conclusion is unsurprising, since most GCs have low masses and that they are sparsely distributed throughout the halo. Because the population of GCs is not expected to change much over the last 5 Gyr, therefore their effect on the Pal 5 stream's density is not expected to be any different.

\section{Discussion and conclusion}
\label{sec:conclusion}

\begin{figure}
\includegraphics[width=0.5\textwidth]{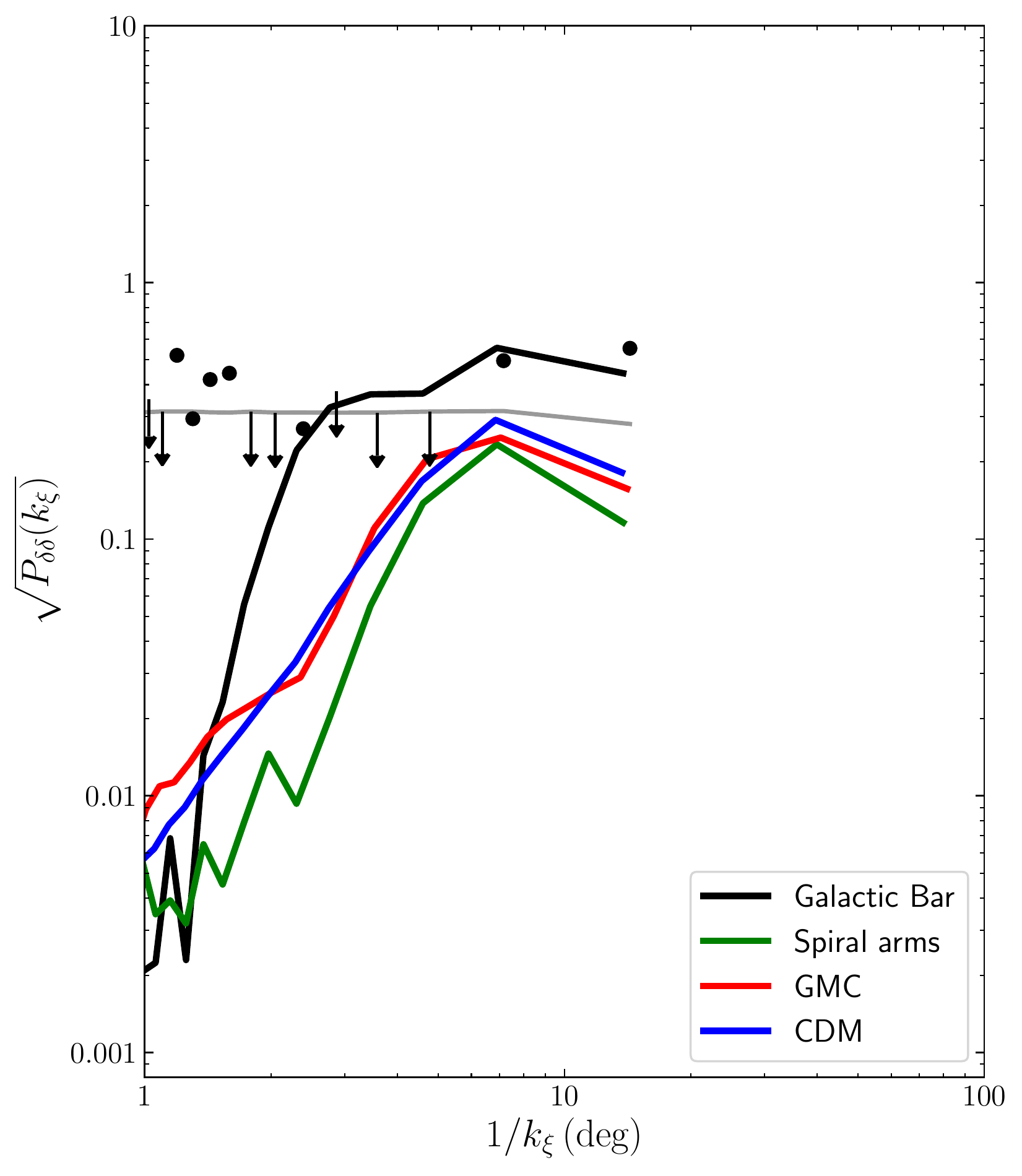}
\caption{Summary of the results of this work. Each curve shows the power spectrum of the mock Pal 5 stream's density as a result of perturbations from the different baryonic structures considered in this paper. The black curve shows the power induced by a 5 Gyr old bar of mass $10^{10} \, \rm{M}_{\odot}$ rotating with a pattern speed 39 $\rm{km \ s^{-1} \ kpc^{-1}}$. The green curve is the power due to four-armed, 3 Gyr old spiral structure whose amplitude corresponds to 1\% of the radial force at the location of the Sun due to the axisymmetric background, and rotating with a pattern speed of 19.5 $\rm{km \ s^{-1} \ kpc^{-1}}$. Both the bar and spiral curves are constructed using $5\times 10^{6}$ points along the stream to minimize the shot noise (the noise is therefore $1/\sqrt{10}$ times lower than in the results in Section \ref{sec:bar_effect} and \ref{sec:spiral_effect}). The red curve is the median power imparted by 40 different realizations of the GMC population on the Pal 5 stream and the blue curve indicates the median power due to CDM subhalo impacts. The black dots are the power and the gray horizontal line is the noise power of the observed Pal 5 stream as computed in ~\protect\citet{Bovy2016a}.}
\label{fig:final_plot}
\end{figure}

In this paper, we have presented an in-depth analysis of the effects of the baryonic structures in our Galaxy on the density of the Pal 5 stream. We considered the effect from the Galactic bar, the spiral arms, and the Galactic population of GMCs and GCs. We examined the effect of each perturber separately by varying their model parameters within limits set by observations and quantified the perturbation imparted to the stream by computing the power spectrum of the stream's density relative to a smooth fit. Figure \ref{fig:final_plot} presents a summary of our findings. In this figure, the density power spectrum of the Pal 5 stream for the four different types of perturbers is shown and compared to the observed power spectrum from \citet{Bovy2016a}. For the bar and spiral structure models we choose a representative example, while for GMCs and dark-matter subhalos we present the median expectation from different realizations of the population (we do not show the GCs, because their power is negligible).

On large scales, where the current observations are dominated by signal rather than noise, the bar, the GMCs, and CDM subhalos can produce power in the density similar to the observed power. Note that Figure \ref{fig:final_plot} shows the median power of the CDM subhalo and GMC impacts, but does not show the dispersion in them; the dispersion in power due to the CDM subhalo impacts is similar to the dispersion in power due to the GMCs that is shown in Figure \ref{fig:GMC_powerdisp}. This implies that constraining the CDM subhalo population using the large-scale power, as was done by \citet{Bovy2016a}, is complicated. Because the exact parameters of the bar, spiral structure, and the GMC population are still uncertain, it is difficult to predict exactly how much power they induce. But for the bar models in particular, the generic prediction from our modeling in Section \ref{sec:bar} is that much power is induced on large scales and the bar must therefore contribute much of the power. Thus, little room is left for dark-matter subhalos to contribute to the power on large scales. 

On small scales, the effect of the bar and spiral structure diminishes strongly and on $\approx 1^\circ$ scales, they drop below the predicted power from dark-matter substructure. The power due to impacts with the GMC population is similar to that due to CDM subhalo. GMC impacts are difficult to distinguish from those from dark matter subhalos. The only difference is that (a) they occur when the stream passes through the disk, while dark matter subhalo impacts occur while the stream is in the halo, and (b) GMCs are about five times more compact. While these differences may in principle be used to distinguish the GMCs and dark matter subhalos, in practice \citet{Bovy2016a} showed that the exact time of impact matters little after a few orbits and that the effect of the concentration on the power spectrum is degenerate with the number of perturbers. Thus, the large effect of GMCs on the Pal 5 stream's density is a largely insurmountable issue when using the stream to constrain the dark matter subhalo population.

Based on  the analyses in this paper, it is clear that the Pal 5 stream is heavily affected by the baryonic structures in our Galaxy. While better constraints on the properties of the bar and spiral structure in the Milky Way might allow us to account for their deterministic effect, the stochastic nature of the Pal 5  stream's interaction with the Galactic population of GMCs severely limits its usefulness for constraining the dark matter subhalo population in the inner Milky Way. Figure \ref{fig:final_plot} demonstrates that better observations of Pal 5's density should uncover large small-scale density fluctuations like those in Figure \ref{fig:GMC_dens} that are due to GMC impacts. This could provide a useful constraint on the total population of high-mass ($M \gtrsim10^5\,\rm{M}_\odot$) GMCs and on the evolution of the GMC mass function. That we can measure the properties of the high-mass GMC population using the Pal 5 stream will also be useful in determining the, hopefully lesser, effect of GMCs on other cold streams.

For constraining dark matter substructure, it is necessary to consider other cold streams. A prime example is the GD-1 stream \citep{Grillmair2006}, which may be a more suitable candidate, because it is situated farther away from the Galactic center with a perigalacticon of $\approx 14$ kpc. In addition, GD-1 is on a retrograde orbit with respect to the disk and therefore the effect of the bar, spiral structure, and GMC population is expected to be minimal \citep[e.g.,][]{Amorisco2016,Erkal2017}. However, baryonic effects may still play a minor role and the methodology in this paper could be applied with few changes to determine their effect on GD-1 or any other stream.

\section*{Acknowledgements}

We thank Gianfranco Bertone,and Denis Erkal for useful discussions and comments. We thank the anonymous referee whose comments have greatly improved this manuscript. NB acknowledges the support of the D-ITP consortium, a programme of the Netherlands Organization for Scientific Research (NWO) that is funded by the Dutch Ministry of Education, Culture and Science (OCW). JB acknowledges the support of the Natural Sciences and Engineering Research Council of Canada (NSERC), funding reference number RGPIN-2015-05235, and from an Alfred P. Sloan Fellowship.




\bibliographystyle{mnras}
\bibliography{refs} 








\bsp	
\label{lastpage}
	\end{document}